\documentclass[preprint,12pt]{elsarticle}

\usepackage{tikz}
\usepackage{multirow}
\usepackage{booktabs}
\usepackage{placeins}
\usepackage{adjustbox}
\usepackage{amsfonts}
\usepackage{tikz-network}
\usepackage{url}
\usepackage{hyperref}
\usepackage{caption}
\usepackage{subcaption}

\usepackage{amssymb}
\usepackage{amsmath}

\journal{Physica A}

\begin{document}

\begin{frontmatter}

\title{Information Flow in the FTX Bankruptcy: A Network Approach}

\author[label1]{Riccardo De Blasis}

\affiliation[label1]{organization={Department of Management, Marche Polytechnic University},
            addressline={Piazzale R. Martelli 8}, 
            city={Ancona},
            postcode={60121}, 
            state={AN},
            country={Italy}}

\author[label2,label3]{Luca Galati\corref{cor1}}
\ead{luca.galati@unibo.it}
\ead[https://www.unibo.it/sitoweb/luca.galati/en]{personal page}

\affiliation[label2]{organization={Department of Management, Alma Mater Studiorum - University of Bologna},
            addressline={Via Capo di Lucca 34}, 
            city={Bologna},
            postcode={40126}, 
            state={BO},
            country={Italy}}

\affiliation[label3]{organization={School of Business, University of Wollongong},
            addressline={Northfield Ave}, 
            city={Wollongong},
            postcode={2500}, 
            state={NSW},
            country={Australia}}
\cortext[cor1]{Corresponding author.}

\author[label5]{Rosanna Grassi}

\affiliation[label5]{organization={Department of Statistics and Quantitative Methods, University of Milano-Bicocca},
            addressline={Piazza dell’Ateneo Nuovo 1}, 
            city={Milano},
            postcode={20126}, 
            state={MI},
            country={Italy}}

\author[label6]{Giorgio Rizzini}

\affiliation[label6]{organization={Faculty of Sciences, Scuola Normale Superiore},
            addressline={Piazza dei Cavalieri 7}, 
            city={Pisa},
            postcode={56126}, 
            state={PI},
            country={Italy}}

\begin{abstract}

This paper investigates the cryptocurrency network of the FTX exchange during the collapse of its native token, FTT, to understand how network structures adapt to significant financial disruptions, by exploiting vertex centrality measures. 
Using proprietary data on the transactional relationships between various cryptocurrencies, we construct the filtered correlation matrix to identify the most significant relations in the FTX and Binance markets. By using suitable centrality measures - closeness and information centrality - we assess network stability during FTX's bankruptcy. The findings document the appropriateness of such vertex centralities in understanding the resilience and vulnerabilities of financial networks. 
By tracking the changes in centrality values before and during the FTX crisis, this study provides useful insights into the structural dynamics of the cryptocurrency market. Results reveal how different cryptocurrencies experienced shifts in their network roles due to the crisis. 
Moreover, our findings highlight the interconnectedness of cryptocurrency markets and how the failure of a single entity can lead to widespread repercussions that destabilize other nodes of the network. 

\end{abstract}


\begin{highlights}
\item Centrality measures trace network structure dynamics in FTX's crisis
\item Network metrics reveal cryptocurrencies' vulnerabilities 
\item More sophisticated altcoins show their centrality over popular cryptocurrencies
\item Network analysis is crucial for risk assessment and crisis management in DeFi
\end{highlights}


\end{frontmatter}

\section{Introduction}
\label{sec:intro}

In recent years, the cryptocurrency market has gained a lot of attention from academics and practitioners \citep[see, e.g.,][]{guo2020}. According to \cite{fang2022}, since 2017, the number of publications on cryptocurrencies has rapidly increased, with more than 140 papers published only in 2021. Almost half of this research focuses on the prediction of returns and volatility, while 20\% studies the relationship between pairs and portfolios, and only about 7\% of researchers are interested in bubbles and extreme conditions of this emerging market. Moreover, the majority of this research (almost 70\%) employs statistical and machine learning methods, and only a partial focus is reserved for the study of the cryptocurrency market from a network perspective. As such, this study aims to fill these critical gaps in the literature by examining the network market structure of the bankrupted cryptocurrency exchange FTX during the collapse of its token, FTT.

Cryptocurrency markets, as traditional financial markets, are characterized by a high number and possibly different types of interactions among various market participants. Due to the intricate nature of such markets, exploiting complex network tools can be highly effective in revealing hidden attributes and determining the importance of each cryptocurrency within the market as a whole.
In particular, by the use of centrality measures, it is possible to assess how central a cryptocurrency is. This allows for the identification of the relevant cryptocurrencies not only on the basis of the connections with other cryptocurrencies in the market but also on the information that each can transmit to others. Notably, the use of these measures is crucial to reveal the signs of the arrival of a crisis on the market.

In the complex network literature, cryptocurrency markets have been analyzed following two main approaches. Part of the literature directly analyses the blockchain public data constructing the cryptocurrency transaction networks, which are the largest real-world networks with publicly accessible data (see, e.g., \cite{wu2021}). Conversely, other studies concentrate on the network constructed from the analysis of the price series. Among possible approaches to construct networks from time series (see \cite{zou2019} for a review), a particularly powerful method involves the construction of the correlation matrix: where nodes represent the assets, and weighted edges capture the relationship among couples of nodes by means of the correlation coefficient. However, this approach constructs a full matrix that does not allow the identification of the informative structure of the network and the filtering of relevant information. In the literature, there are two main approaches to reducing information's redundancy in the correlation matrix and building sparse networks containing only relevant edges: (i) the minimum spanning tree (MST) and (ii) the planar maximally filtered graph (PMFG). 

The approach proposed in the seminal work  \cite{mantegna1999} belongs to the first stream of research. The author proposes a method filtering the most important $n-1$ links from a $n \times n$ correlation matrix to construct the MST by introducing an ultrametric that transforms correlations into distances. \cite{onnela2003} follows a similar approach, first building the network of stocks from the New York Stock Exchange (NYSE) and then extracting the MST from the distance matrix. 
Regarding the second filtering approach, the authors in \cite{tumminello2005} extend the methodology in \cite{mantegna1999} proposing a heuristic algorithm to construct the Planar Maximally Filtered Graph (PMFG) 
filtering the correlation matrix. \cite{tumminello2007} analyzes the topological features of a class of PFMG networks from the returns of the 300 most capitalized stocks traded at NYSE during the period 2001–2003 at different time horizons. Recently, \cite{massara2016} proposes a new efficient algorithm to filter the correlation matrix based on triangulation called Triangulated Maximally Filtered Graph (TMFG).

Among the network applications to cryptocurrencies, \cite{pilar2018} examines the MST network built from the correlations of daily returns of 16 cryptocurrencies following the approach in \cite{mantegna1999}.
From a static analysis of the network, the authors identify the Ethereum currency in a central position of the MST, then as the benchmark within the market, leaving Bitcoin in a peripheral position. 
From a dynamic network perspective, \cite{scagliarini2022} analyzes the effects of information flows in cryptocurrency markets built from the Granger causality\footnote{Granger causality networks are graphs in which two nodes are connected if one of them causes in Granger meaning (see \cite{granger1969}) the other one.} among weekly log-returns and find a quite stable network structure over time. On the contrary, \cite{nie2022} examines the stability of the PMFG cryptocurrency network around critical events using a function of neighbours' influence strength. The author shows that critical events lead to significant changes in the structure of the network from stability to fragility, going back to stability once the critical time has passed. \cite{liao_interconnections_2024}, instead, explores the dynamic downside risk among digital financial assets (cryptocurrencies, DeFi tokens, and NFTs) and traditional financial assets (stock indices and commodities) by constructing a daily network based on the CoVaR measure. They document significant tail risk spillovers, bidirectional between digital assets and commodities, and unidirectional from stocks to digital assets and from commodities to stocks. 

To this extent, it is worth noting that a major critical event has occurred in the cryptocurrency industry, the collapse of the FTX market in November 2022. This event has been analyzed in the literature from different viewpoints. From a \textit{market 
microstructure perspective}, \cite{galati_market_2024} contributes to understanding the systemic implications of major disruptions in the cryptocurrency markets. In particular, the authors examine how the halt in withdrawals at a major exchange affects market liquidity, traders' behavior, and asset pricing dynamics. Their findings indicate significant liquidity deterioration and effectively highlight the critical vulnerabilities within the cryptocurrency market infrastructure, especially under stress scenarios like bankruptcy and operational disruptions. Similar to \cite{de_blasis_intelligent_2023} for the collapse of the Terra-Luna token and \cite{galati_silicon_2024} for the Silicon Valley bank bankruptcy, \cite{galati_financial_2024} uses a BEKK Generalized AutoRegressive Conditional Heteroskedasticity (GARCH) model to examine the intraday volatility spillover effects of the FTX collapse to other cryptocurrency exchanges. The authors found evidence consistent with their proposition, and also examined the information cascade effects of other cryptocurrency assets on FTX when nearly all withdrawals were prohibited. Abnormal returns for major assets indicated a flight to safety from less to more authoritative digital assets. \cite{esparcia_assessing_2024} further corroborates these findings, showing that the FTX collapse increased overall intraday volatility in the cryptocurrency markets, with stablecoins being the most affected. Their use of the Multiplicative Component GARCH (MCGARCH) model and Time-Varying Parameter VAR model (TVP-VAR) methodology revealed FTT's role as a primary volatility transmitter, particularly impacting stablecoins like USD Coin and altcoins like Polygon, thus highlighting the interconnectedness and systemic risk in the cryptocurrency market. 

The recent scholarly examinations of the FTX collapse shed light on its broader implications as well. \cite{jalan_systemic_2023} finds surprisingly stable systemic risks and liquidity despite the collapse, suggesting inherent market resilience against such shocks. \cite{akyildirim_understanding_2023} identifies specific contagion effects propagated by FTX-associated tokens like FTT and Serum, which significantly influenced related financial products. The study  \cite{yousaf_reputational_2023} reveals that while cryptocurrencies like Bitcoin, Ethereum, and Binance experienced significant negative returns due to the FTX collapse, traditional asset markets remained unaffected, indicating limited contagion. In a similar vein, \cite{yousaf_what_2023} finds that the FTX collapse significantly impacted cryptocurrency markets but not traditional financial assets, showing traditional investors' indifference to cryptocurrency fluctuations during bear markets. \cite{conlon_collapse_2023} examines how market sentiment and investor behavior changed in response to news releases related to FTX, while  \cite{conlon_contagion_2024} argues for enhanced regulatory oversight in crypto markets to mitigate financial instability risks. 

From a \textit{network perspective}, the FTX collapse has been studied by \cite{wang2024}, which examines the cross-exchange risk in a high-frequency setting employing a Multivariate Heterogeneous AutoRegression (MHAR) model to build the network. The network is constructed by filtering the correlations which exceed in absolute value a fixed threshold.
The authors find that FTX bankruptcy triggered a chain effect among exchanges, i.e., increased partial correlations with other exchanges, and highlighted the spillover risk and its persistent effect across centralized exchanges. Additionally, \cite{vidal-tomas_ftxs_2023} investigates the causes and consequences of FTX’s failure. Their results reveal that leveraging and misuse of its native token, FTT, exacerbated FTX's financial fragility highlighting the Terra-Luna collapse as a pivotal trigger event. The study exploits the TMFG to model the evolutionary dependency structures of 199 cryptocurrencies, showing the systemic impact of Binance's actions on FTX's downfall and emphasizing the trend toward centralization in the crypto market.

To the best of our knowledge, this study is the first that combines market microstructure theory and network analysis to investigate the FTX collapse. It provides insights into the evolving patterns of suitable centrality measures throughout the crisis of one the largest cryptocurrency centralized exchanges, FTX, and unveils the intricate relationships between market behaviors and investor responses, which, in turn, modify the cryptocurrency network structures before and during financial disruptions. In addition, our study provides suitable measures for assessing network resilience and stability. In particular, it highlights the crucial roles that different cryptocurrencies play in sustaining the informational architecture of the market during crises. 

The results offer a comprehensive picture of resilience and vulnerability inherent in financial networks, especially within such significant markets as FTX and Binance, the industry leader (see, e.g., \citep{galati_exchange_2024}).
This also hints at potential strategic pivot points for stakeholders within the cryptocurrency ecosystem. Indeed, understanding the centrality dynamics can equip market participants, analysts, and regulators with deeper insights into the critical nodes within the networks. This is particularly useful because it potentially drives investment decisions, risk assessments, and regulatory considerations. This study, thus, contributes to a better comprehension of the structural complexities within digital asset markets, revealing the intricate interplay of connections, resilience, and influence.

The rest of the paper proceeds as follows. Section \ref{sec:method} details the methodology adopted through a theoretical background on network theory, centrality measures and filtered graphs. A toy example is proposed to show the appropriateness of the proposed centrality measures in cryptocurrency networks. Section \ref{sec:results} describes the data and the empirical findings. Section \ref{sec:conclusion} concludes.

\section{Methodology}
\label{sec:method}

\subsection{Preliminary definitions on network theory}
\label{sec:graph}

In this section, we briefly review some mathematical network definitions.
A network is formally represented by a graph $G=(V,E)$, where $V$ is the set of $n$ nodes (or vertices) and a set $E$ of $m$ edges (or links). Two nodes $i$ and $j$ are adjacent if there exists an edge connecting them, i.e., if $(i,j) \in E$. $G$ is undirected if $(j,i)\in E$ whenever $(i,j)\in E$. $G$ is simple if loops -- a link joining a vertex to itself -- and multiple edges --  more edges incident to the same pair of vertices -- are not allowed. In this paper, we assume that graphs are undirected and simple. A subgraph $G'=(V',E')$ of $G$ is a graph such that $V' \subseteq V$ and $E' \subseteq E$. The degree $d_i$ of a node $i$ is the number of links incident to it. A complete graph $K_n$ is a graph of $n$ nodes such that each node degree is equal to $n-1$.  A graph is weighted if a non-negative real number $w_{ij}$ is associated with each edge $(i,j)$ of $G$. 

The adjacency relations can be represented by a $n-$square matrix $\mathbf{A}$, the \textit{adjacency matrix}, whose elements $a_{ij}$ are equal to 1 if $(i,j) \in E$, and $0$ otherwise. As graphs are simple and undirected, the diagonal entries of $\mathbf{A}$ are null, and $\mathbf{A}$ is symmetric. The adjacency relations for a weighted graph are represented by a $n-$square matrix $\mathbf{W}$, with elements $w_{ij}$ if the weighted link $(i,j)\in E$, and $0$ otherwise. 
An $i-j$ path is a sequence of distinct adjacent vertices from $i$ to $j$. The distance $d(i,j)$ between $i$ and $j$ is the length of the shortest path joining them when such a path exists, and it is set to $+\infty$ otherwise. A cycle of length $k$ is a path of $k$ edges in which the first and the last vertices coincide.

A graph $G$ is connected if there is a path between every pair of vertices. A connected component of $G$ is a maximal set of nodes such that each pair of nodes is connected by a path. A connected graph has exactly one connected component.
For a more detailed treatment of graphs and networks, we refer the reader, for instance, to \cite{Harary} and \cite{Estrada2012}.
Since the mathematical object underlying the network is the graph, in the rest of the text, we will use the words graph and network interchangeably.

Finally, we introduce the concepts of planar graphs and maximally planar graphs. A planar graph is a graph that can be drawn in such a way that no edges cross each other, see \cite{west2001}. A necessary but not sufficient condition for a graph $G$ to be planar is that $m \le 3n-6$ for $n\ge 3$. Notice that this condition implies that planar graphs are sparse networks characterized by a number of edges of order $O(n)$.

Planar graphs are useful in extracting relevant information from a complex database represented by a weighted graph. 
Indeed, in such a complex structure, it can be meaningful to filter data by unveiling the most significant information. This can be done by searching for the largest possible subgraph satisfying some topological constraints.
In the related literature about networks, this problem is known as the Weighted Maximal Planar Graph problem (see \cite{Nishizeki2008}). In the financial context, a first attempt to solve such a problem is presented in \cite{mantegna1999} in which the information's filtering procedure is performed by extracting the Minimum Spanning Tree (MST), 
that is a subgraph on $n$ nodes, connected and without cycles, with $n - 1$ edges, such that the sum of all edge weights is minimized.

A more sophisticated procedure to extract relevant information consists of constructing the Planar Maximally Filtered Graph (PMFG), which belongs to the class of Information Filtering Networks (see \cite{tumminello2005,aste2005}). Specifically, given 
the weighted adjacency matrix of a complete graph $K_n$, the authors propose an algorithm to extract the maximal planar subgraph, with $n$ vertices and $3(n-2)$ edges, ensuring the highest sum of edge weights.  
The construction of PMFG is based on the following procedure: all the edges of the initial dense network are sorted in non-increasing order, and one edge per time is added to the PMFG. Edges that violate the planarity constraint are discarded. Edges are added until the PMFG has exactly $3(n-2)$ edges. Recently \cite{massara2016} proposed a new algorithm to determine the solution of the Weighted Maximal Planar Graph problem. The resulting subgraph is based on a triangulation obtained by maximizing a score function linked to the information withheld by the starting network.

\subsection{Centrality measures based on paths}
\label{sec:centrality}

In network theory, centrality is one of the key issues. Broadly speaking, any element of the network  (nodes, edges or groups of nodes) plays a role with respect to the global network structure. However, to assess their relevance in terms of connections, the most studied aspect of centrality is the assignment of a score to the vertices.

The degree centrality of a node $i$ is the most intuitive centrality measure as it counts the number of neighbours of $i$, and it is formally represented by the normalized degree $d_i=\frac{1}{n(n-1)} \sum_{j=1}^{n} a_{ij}$.

With reference to the information flow, an interesting class of centrality measures is those based on the concept of distance in the network, namely paths between pairs of nodes. A relevant role in this framework is certainly played by the closeness centrality (see \cite{Freeman1978}). This measure is based on the length of the paths from a node $i$ to all other nodes in the network, and formally, it is defined as the reciprocal of the sum of the distance between $i$ and all other nodes (multiplied by $n-1$ to obtain a normalized measure):

 \begin{equation}
 c(i)=\frac{n-1}{\sum_j d(i,j)}.
 \label{eq:closeness_theory}
 \end{equation}

A meaningful weighted version has been proposed by \cite{Opsahl2010}, based on the idea of weighted shortest paths. The identification of the shortest paths is quite simple for unweighted networks. 
Indeed, 
the geodesic distance between two nodes $i$ and $j$ is the length of the path with the minimum number of edges connecting $i$ and $j$. The matter is more complicated when links are weighted. 
The problem of identifying the weighted shortest path has been analyzed in many papers, e.g., see \cite{schrijver2012} among others, 
most of them based on Dijkstra's algorithm, where weights on the links are interpreted as costs of transmission (\cite{Dijkstra1959}). 
In the milestone papers of \cite{newman2001} and \cite{brandes2001} the authors propose to invert the link's weight before applying the Dijkstra's algorithm. 
Indeed, a low weight of the link makes the passage through it more costly than passing through a link with a high weight.\footnote{Indeed, given two links weights $w_{ij}$ and $w_{hk}$ such that $w_{ij}>w_{hk}$ then $\frac{1}{w_{ij}}< \frac{1}{w_{hk}}$.} In such a framework, the weighted shortest path distance between nodes $i$ and $j$ is defined as
\begin{equation}
    d^w(i,j)=\min\left(\frac{1}{w_{ih}}+\dots+\frac{1}{w_{hj}}\right),
    \label{eq:distance_opshal}
\end{equation}
and the normalized weighted closeness centrality measure as
\begin{equation}
    c^w(i) = \frac{n-1}{\sum_j d^w(i,j)}.
    \label{weight_closeness}
\end{equation}



Another centrality path-based measure is information centrality. This measure is based on the concept of efficiency introduced by \cite{latora2001} to assess how much nodes in a network exchange information. The authors assume that the information between two nodes $i$ and $j$ is spread through one of the possible weighted shortest paths between the involved nodes. In this way, the efficiency $\epsilon^w(i,j)$ 
is defined as the reciprocal of the shortest distance. Indeed, the higher the geodesic distance between $i$ and $j$, the lower their efficiency in the information's transmission. The efficiency $\epsilon^w(i,j)$ between $i$ and $j$ is therefore defined as 
\begin{equation}
    \epsilon^w(i,j) = \frac{1}{d^w(i,j)},
    \label{eq:efficiency_link}
\end{equation}
so that if $d^w(i,j)= + \infty$, i.e.\ there is no a shortest path between $i$ and $j$, then $\epsilon^w(i,j) = 0$ and therefore no information can travel between those nodes. 
From \eqref{eq:efficiency_link}, the efficiency of a graph $G$ 
naturally arises:
\begin{equation}
    \varepsilon^w(G) = \frac{1}{n(n-1)} \sum_{\substack{i,j\in V\\ i \ne j}} \epsilon^w(i,j).
    \label{eq:efficiency}
\end{equation}


Giving the previous definition, the information centrality of a node $i$ is defined as (\cite{latora2007}):
\begin{equation}
c_I(i) = 1 - \frac{\varepsilon^w(G’)}{\varepsilon^w(G)},
\label{eq:inf_centrality}
\end{equation}
where $G'$ is the subgraph obtained by removing from $G$ all the edges incident to node $i$, i.e., broadly speaking, the node $i$ in $G’$ is an isolated node.
This measure is suitable to capture how a network reacts when a node $i$ is deactivated, i.e., when node $i$ cuts all its links. Thus, it measures the relative reduction in the
network efficiency after the node $i$'s removal. Information centrality ranges in $[0,1]$: if $c_I(i) = 0$ it means that the removal of node $i$ does not affect the efficiency of $G'$, i.e.,\ $\varepsilon^w(G)=\varepsilon^w(G’)$ while $c_I(i) = 1$ means that $G'$ is an empty graph, i.e.\ the edge set of $G'$ is empty. \\
Therefore, among the centrality measures described in this section, the latter is the most significant in terms of information transmission. Moreover, in Section \ref{sec:results2}, we use a global centrality measure by  averaging nodes' information centrality 
   $ c_I(G) = \frac{1}{n} \sum_{i} c_I(i). $

\subsection{A toy example}
\label{sec:example}

In this section, by means of a simple example, we show the appropriateness of the proposed centrality measures in capturing the role of each node in the transmission of information within the network.\ The network $G$ is plotted in Figure \ref{fig:example_network} whose weighted adjacency matrix is:

\begin{figure}
     \centering
     \begin{subfigure}[b]{0.3\textwidth}
         \begin{tikzpicture}
		\Vertex[label=a,x = 15, y = 1]{A} \Vertex[label=b, x=15, y = -1]{B} \Vertex[label = c, x=17,y=0]{C} \Vertex[label = d, x=13,y=0]{D} \Vertex[label = e, x=15,y=3]{E}
		\Edge[label=0.5,position={above right=2mm},distance=.7](A)(B)
		\Edge[label=0.2,position=above](A)(C)
		\Edge[label=0.3,position=above](A)(D)
		\Edge[label=0.2,position=above](B)(D)
		\Edge[label=0.1,position=above](B)(C)
		\Edge[label=0.4,position={above right=2mm},distance=.3](A)(E)
	\end{tikzpicture}
    \caption{Network $G$ with $\varepsilon^w(G) = 0.24$.}
    \label{fig:example_network}
     \end{subfigure}
     \hspace{20mm}
     \begin{subfigure}[b]{0.3\textwidth}
       \begin{tikzpicture}
		\Vertex[label=a,x = 15, y = 1]{A} \Vertex[label=b, x=15, y = -1]{B} \Vertex[label = c, x=17,y=0]{C} \Vertex[label = d, x=13,y=0]{D} \Vertex[label = e, x=15,y=3]{E}
		\Edge[label=0.2,position=above](B)(D)
		\Edge[label=0.1,position=above](B)(C)
	\end{tikzpicture}
 \caption{Network $G'$ with $\varepsilon^w(G') = 0.04$.}
\label{fig:example_network_after_node_a_elimination}     
\end{subfigure}
        \caption{Networks $G$ and $G'$ representation.}
\end{figure}
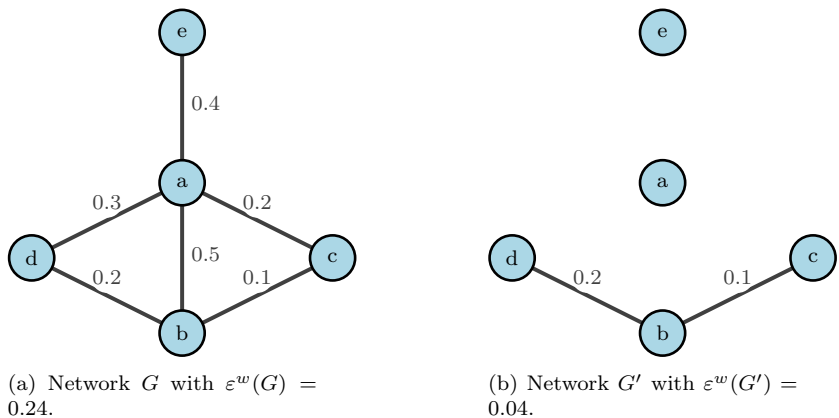






\begin{equation}
   \mathbf{W} = \begin{bmatrix}
         0.0 & 0.5 & 0.2 & 0.3 & 0.4 \\
        0.5 & 0.0 & 0.1 & 0.2 & 0.0 \\
         0.2 & 0.1 & 0.0 & 0.0 & 0.0 \\
        0.3 & 0.2 & 0.0 & 0.0 & 0.0 \\
        0.4 & 0.0 & 0.0 & 0.0 & 0.0 \\
    \end{bmatrix}.
    \label{W_toy_example}
\end{equation}

We first compute the centrality measures degree and closeness; then the information centrality of $G$ by using Formula \eqref{eq:inf_centrality}. 
Notice that the closeness has been computed applying Formula \eqref{weight_closeness}, which is based on the reciprocal of the link's weight.  \\
Table \ref{tab:degree_example} reports node rankings based on the three centrality measures. We observe that the node $a$ is the one with the highest information centrality, highlighting its strong presence in paths connecting nodes and, therefore, confirming its central role in spreading information in the network.\\ Consider in the network $G$ the four weighted shortest paths connecting nodes $d$ and $c$ (i.e.,\ $d-a-c, d-b-c, d-a-b-c, d-b-a-c$). Among these paths, the shortest one, i.e.\ the one with the smallest cost, is $b-a-c$ with $d^w(d,c)= 8.33$. We notice that the node $a$ is present in three out of four paths connecting $d$ and $c$, in line with its central role in the network. After the node $a$'s removal (graph $G'$ in Figure \ref{fig:example_network_after_node_a_elimination}), there exists only one path connecting $d$ and $c$ with a cost of $15$. This result confirms that the higher the information centrality of the deactivated node, the higher the cost of information transmission in the network. Table \ref{tab:shortpaths} reports all the weighted paths with their costs both in $G$ (Panel A) and $G'$ (Panel B). \\
Similar considerations can be done by deactivating one node of $G$ per time and the impact in terms of graph efficiency is shown in Table \ref{tab:efficiency}. The results confirm the predominant role of node $a$ in the network: indeed, the deepest drop in the network efficiency is observed when the node $a$ is deactivated, while the less deep efficiency is associated with the deactivation of the node $e$.

\begin{table}
    \centering
    \begin{tabular}{cc|cc|cc}
    \toprule
    Ranking & $d_i$  &  Ranking & $c^w(i)$ &   Ranking & $c_I(i)$ \\
    \hline \hline
   $a$  & 1.00 &   $a$ & 0.28 &  $a$ & 0.85 \\
   $b$  & 0.75 &  $b$ & 0.19 &   $b$ & 0.45 \\
   $c$ & 0.50 &   $e$ & 0.17 &   $e$ & 0.39 \\
    $d$ & 0.50 &   $d$ & 0.16 &   $d$ & 0.33 \\
    $e$ & 0.25 &   $c$ & 0.13 &   $c$ & 0.25 \\
    \bottomrule
    \end{tabular}
    \caption{Node rankings based on degree, closeness, and information centralities of the network $G$ depicted in Figure \ref{fig:example_network}.}
    \label{tab:degree_example}
\end{table}

\begin{table}
    \centering
    \begin{tabular}{lllr}
    \toprule
    \midrule
    \multicolumn{4}{l}{Panel A: shortest path in $G$} \\
    $i$ & $j$ & Shortest path & $d^w(i,j)$ \\
    \midrule
   $ b$ & $a$ & $b-a$ & 2.00 \\
    $e$ & $a$ & $e-a$ & 2.50 \\
    $d$ & $a$ & $d-a$ & 3.33 \\
    $c$ & $a$ & $c-a$ & 5.00 \\
    $e$ & $b$ & $e-a-b$ & 4.50 \\
    $d$ & $b$ & $d-b$ & 5.00 \\
   $ e$ & $d$ & $e-a-d$ & 5.83 \\
   $ c$ & $b$ & $c-a-b$ & 7.00 \\
   $ e$ & $c$ & $e-a-c$ & 7.50 \\
    $d$ & $c$ & $d-a-c$ & 8.33 \\
    \midrule
    \multicolumn{4}{l}{Panel B: shortest paths in $G'$, i.e.\ when node $a$ is deactivated} \\
        \midrule
        $i$ & $j$ & Shortest path & $d^w(i,j)$ \\
    \midrule
    $d$ & $b$ & $d-b$ & 5.00 \\
    $c$ & $b$ & $c-b$ & 10.00 \\
    $c$ & $d$ & $c-b-d$ & 15.00 \\
    \bottomrule
    \end{tabular}
    \caption{Shortest paths before and after eliminating all the edges incident to node $a$.}
    \label{tab:shortpaths}
\end{table}

\begin{table}
    \centering
    \begin{tabular}{lr}
    \toprule
    Deactivated node & $e^w(G')$ \\
    \midrule
    $a$ & $0.04$ \\
    $b$ & $0.13$ \\
    $c$ & $0.18$ \\
    $d$ & $0.16$ \\
    $e$ & $0.15$ \\
    \bottomrule
    \end{tabular}
    \caption{Graph efficiency after disconnecting a node.}
    \label{tab:efficiency}
\end{table}

\section{Results}
\label{sec:results}

\subsection{Data}
\label{sec:data}

This study uses cryptocurrency trading price data spanning from October $16^{th}$ to November $16^{th}$, 2022, gathered from FTX,\footnote{\url{https://ftx.com/}} the bankrupted exchange of reference, and Binance,\footnote{\url{https://www.binance.com/}} the major exchange in cryptocurrency markets. However, data from FTX were only available up until November $12^{th}$, 03:28 a.m., due to operational disruptions. Proprietary data were collected from Refinitiv,\footnote{\url{https://www.lseg.com/en}} a London Stock Exchange Group (LSEG) business, and sourced from the DataScope Select database.\footnote{\url{https://www.lseg.com/en/data-analytics/products/datascope-select-data-delivery}}

\begin{figure}
    \centering
    \includegraphics[width=\textwidth]{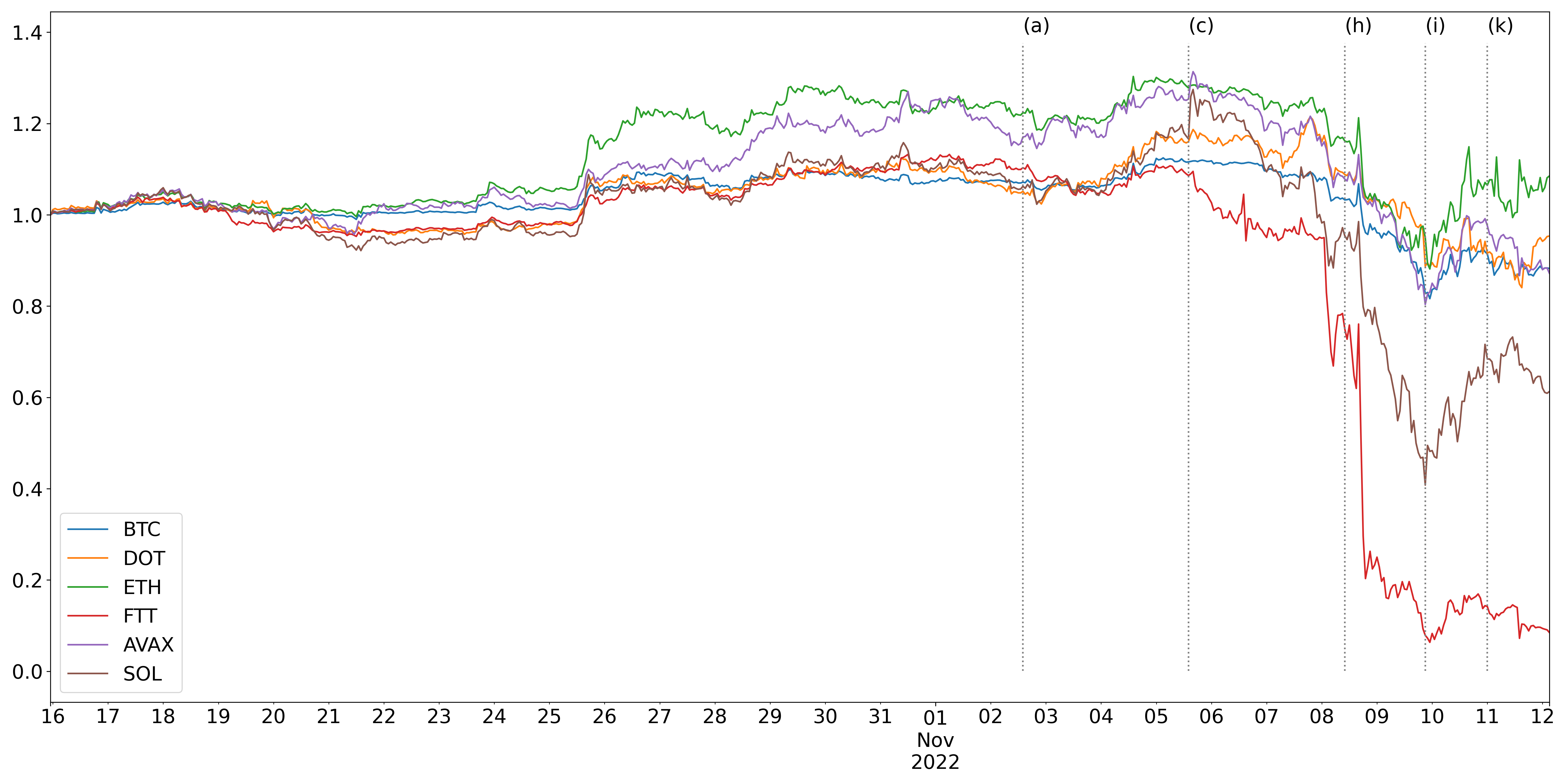}
    \includegraphics[width=\textwidth]{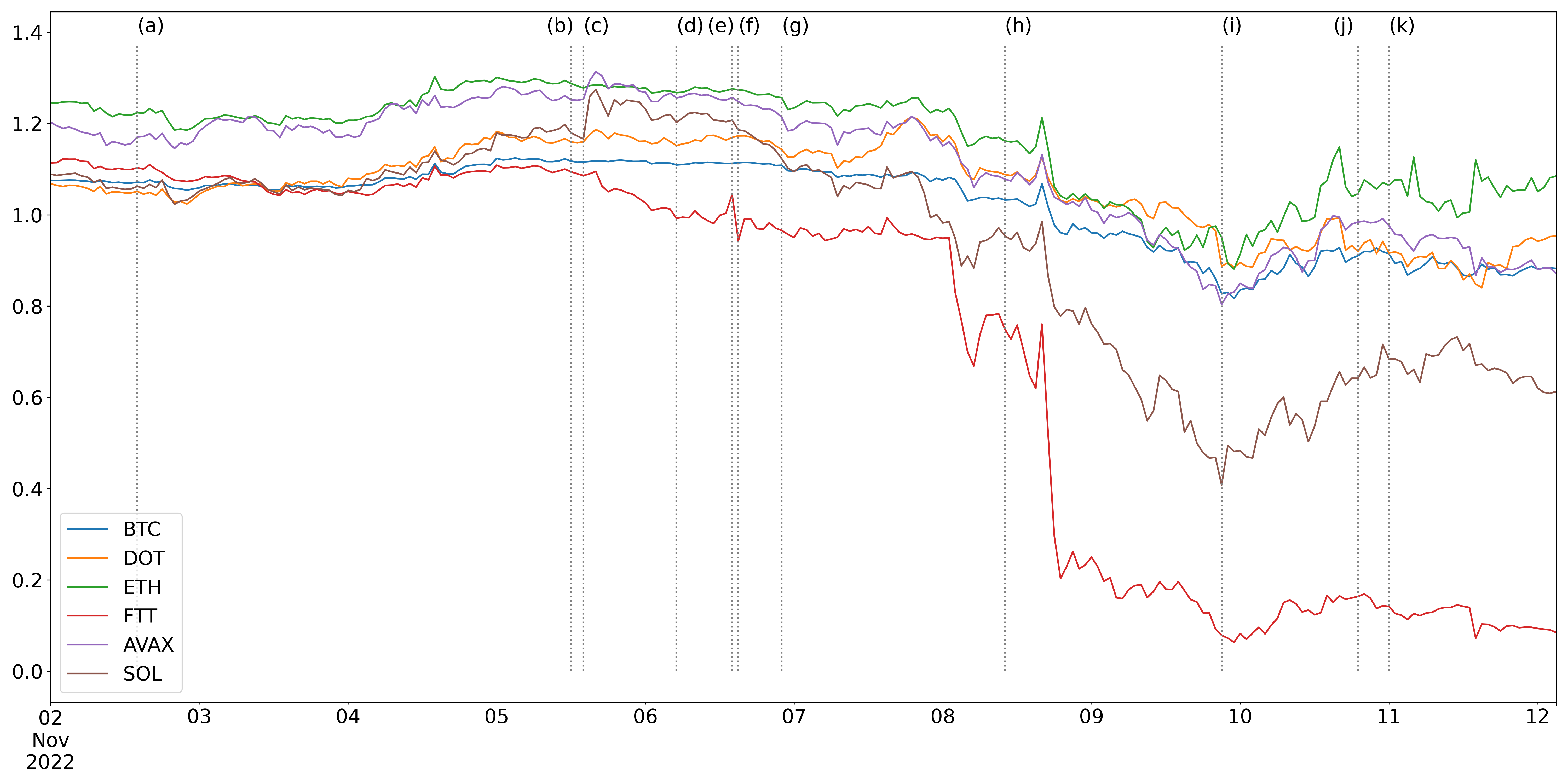}
    \caption{Normalized price series of major cryptocurrencies traded on FTX. Event (c) corresponds to the FTX's disruptions.}
    \label{fig:prices}
\end{figure}

The initial dataset comprises a total of 442 cryptocurrencies on FTX and 344 cryptocurrencies on Binance. In order to maintain focus on purely cryptocurrency movements and exclude external financial influences, several categories of assets were removed from the initial dataset. Specifically, we excluded leveraged tokens (bear and bull), tokens with partial exposure (half), and hedging options (hedge), as well as all fiat currencies, which are not the focus of our study. In addition, to avoid repetition and high correlations between similar crypto-assets, we excluded the cryptocurrency pairs against Tether (USDT) or other digital assets and focused exclusively on trading pairs against the US dollar (USD), being the most liquid and used instrument.\footnote{For Binance, instead, we focus exclusively on trading pairs against the USDT as the exchange does not allow customers to trade against the fiat currency USD.} After these exclusions, the dataset was consolidated into 195 cryptocurrencies traded on FTX and 324 cryptocurrencies traded on Binance.\footnote{The full lists of the cryptocurrencies included in this study are available in the appendix} Nonetheless, as in \cite{galati_market_2024}, the dataset collected from Binance is used as an overall market comparison only, and the focus of this study remains on the actual bankrupted exchange, FTX. The full list of the cryptocurrencies analyzed is in \ref{appendix_list}.

For high-frequency analytical granularity, in the empirical analysis we processed the data to include 1-minute logarithmic returns for each of the remaining cryptocurrencies, computed as $r_{i,t} = \ln\large(\frac{P_{i,t}}{P_{i,t - 1}}\large)$ where $P_{i,t}$ is the price of the cryptocurrency $i$ at time $t$ and $P_{i,t - 1}$ is the price of the cryptocurrency $i$ at time $t-1$. This high-frequency data allows for a detailed analysis of price dynamics and volatility within the specified period across both trading platforms. 

\begin{table}
    \centering
    \begin{tabular}{llp{8cm}}
    \toprule
    Date & Event & Description \\
    \midrule
    2022-11-02 14:44 & (a) & Alameda Research’s (FTX’s sister) balance sheet is possibly compromised and predominantly consists of FTX’s native exchange token, FTT, as reported by CoinDesk. \\
    2022-11-05 12:51 & (b) & Binance’s CEO, Changpeng Zhao (CZ), tweets “Crypto is high risk” \\
    2022-11-05 14:09 & (c) & Binance moves \$585 millions of FTT \\
    2022-11-06 05:59 & (d) & Tweet speculating on proof of FTX insolvency \\
    2022-11-06 14:32 & (e) & Alameda Reserach’s CEO, Caroline Ellison (CE), tweets against insolvency, which was retweeted by FTX’s CEO, Sam Bankman-Fried (SBF)  \\
    2022-11-06 15:47 & (f) & CZ’s tweets on Binance liquidating any remaining FTT \\
    2022-11-06 22:49 & (g) & CZ’s tweet that confirms Binance’s position against FTX on the rumored acquisition \\
    2022-11-07 xx:xx & none & SBF’s tweet that was later deleted against Binance’s supposedly false rumors and defending FTX and its assets \\
    2022-11-08 10:37 & (h) & The last outgoing transaction from FTX on the Ethereum blockchain (i.e., FTX’s halt of funds withdrawals) \\
    2022-11-09 21:00 & (i) & Binance drops out from FTX acquisition \\
    2022-11-10 19:08 & (j) & FTX tweets on reopening some withdrawals for Bahamian customers only  \\
    2022-11-11 00:00 & (k) & FTT token collapsed entirely \\
    2022-11-11 xx:xx & none & FTX filed for Chapter 11 (bankruptcy) \\
    2022-11-12 03:28 & (l) & FTX’s close of business (availability of data from the FTX market) \\
    \bottomrule
    \end{tabular}
    \caption{Timeline of events (UTC time) as in \cite{galati_market_2024}. }
    \label{tab:events}
\end{table}

The visual representations provided in Figure \ref{fig:prices} effectively capture the normalized price trajectories of major cryptocurrencies, including Bitcoin (BTC), Ethereum (ETH), Polkadot (DOT), Solana (SOL), Avalanche (AVAX), and the FTX Token (FTT). These plots are particularly illustrative of the market dynamics in response to the unfolding events during the bankruptcy timeline, as reported in Table \ref{tab:events}. The price data reveal significant volatility, especially visible in the sharp decline of FTT’s value, which directly correlates with the escalated phases of the FTX crisis. The graphical depictions allow for a clear observation of the cascading effects on other cryptocurrencies, underscoring the interconnectedness of the crypto market.

Finally, based on the observation of the price series, we divide the sample into two periods: one preceding the collapse of the FTX’s token, FTT, which arose from the main exchange competitor, Binance, moving a large amount of the token (Event (c) in Table \ref{tab:events}); and one during the collapse period after that event.

\subsection{Network construction}
\label{sec:network_construction}

We construct the network as follows. Firstly, we consider $n= 195$ 
cryptocurrencies and we construct the correlation matrix $\mathbf{C}$ in which the element $c_{ij}$
is the Pearson correlation coefficient between cryptocurrencies $i$ and $j$ returns time series. The correlation matrix 
$\mathbf{C}$ is a full matrix, then its associated weighted graph is a complete $K_n$ graph containing loops. Moreover, the entries can be either positive or negative, as $c_{ij} \in [-1,1]$. According to the literature (\cite{Shirokikh2013,Chen2021}), to overcome these issues we consider the absolute value of the entries, focusing only on the intensity of the assets co-movements, and we set the diagonal elements of $\mathbf{C}$ to zero. We obtain a new non-negative matrix $\tilde{\mathbf{C}}$ such that the associated complete graph becomes simple. At this stage, we need to filter the information in $\tilde{\mathbf{C}}$ to unveil the hidden relevant structure. Therefore, we filter the correlation matrix by constructing the Triangulated Maximally Filtered Graph (TMFG) proposed in \cite{massara2016} which is a computationally efficient algorithm to construct the PMFG (see Sect.\ \ref{sec:graph}). 

\subsection{Descriptive statistics}
\label{sec:descriptive}
In this study, we focus on the major cryptocurrencies (BTC, DOT, ETH, FTT, AVAX, SOL) 
showing the highest logarithmic returns correlations 
in the FTX market during the period preceding the collapse of FTT (Event (c) in Table \ref{tab:events}), which caused the bankruptcy of the FTX exchange, given the fraudulent behaviors with the sister trading company Alameda Research. 
We analyse market conditions in three distinct phases: pre-collapse, during the collapse, and over the entire observed period.
Table \ref{tab:HighCorr} shows the top 20 correlations between the cryptocurrency returns analyzed in the FTX market, highlighting significant variations across the different periods examined. 
The major cryptocurrencies we refer to are those at the top 5 positions 
in the first column of Table \ref{tab:HighCorr}. 

In the pre-collapse phase, higher correlations were observed among major cryptocurrencies, such as a correlation of 0.80 between ETH and BTC. This indicates a closely knit market where the movements of these major currencies were largely synchronous. However, during the collapse, the correlation figures decreased substantially, such as between ETH and BTC dropping to 0.40, reflecting a market dislocation where individual cryptocurrencies responded differently to the crisis. Even stronger evidence is reported for the correlation between SOL and BTC, going from a level of 0.69 in the pre-event to 0.26 within the collapse period, and other cryptocurrency trading pairs. This pattern suggests that the market structure was significantly altered by the event, affecting how asset prices moved in relation to each other. 

\begin{table}
    \centering
    \scriptsize
    \begin{tabular}{lllrlllrlllr}
    \toprule
     & \multicolumn{3}{c}{Pre-collapse} &  & \multicolumn{3}{c}{Collapse} &  & \multicolumn{3}{c}{Full period} \\
    \cmidrule{2-4} \cmidrule{6-8} \cmidrule{10-12}
    1 & ETH & BTC & 0.80 &  & ETH & BTC & 0.40 &  & ETH & BTC & 0.47 \\
    2 & SOL & BTC & 0.69 &  & DOGE & BTC & 0.36 &  & BTC & AVAX & 0.37 \\
    3 & SOL & ETH & 0.68 &  & XRP & BTC & 0.33 &  & XRP & BTC & 0.37 \\
    4 & FTT & BTC & 0.66 &  & DOGE & AVAX & 0.32 &  & DOGE & BTC & 0.33 \\
    5 & SOL & AVAX & 0.65 &  & BTC & AVAX & 0.31 &  & DOGE & AVAX & 0.31 \\
    6 & FTT & ETH & 0.64 &  & DOGE & CHZ & 0.29 &  & ETH & AVAX & 0.30 \\
    7 & SOL & DOT & 0.63 &  & XRP & DOGE & 0.27 &  & DOT & BTC & 0.30 \\
    8 & SHIB & KSHIB & 0.63 &  & ETH & DOGE & 0.27 &  & DOT & AVAX & 0.29 \\
    9 & BTC & AVAX & 0.61 &  & SOL & BTC & 0.26 &  & SOL & BTC & 0.29 \\
    10 & DOT & AVAX & 0.61 &  & SOL & DOGE & 0.25 &  & XRP & ETH & 0.29 \\
    11 & DOT & BTC & 0.61 &  & CHZ & AVAX & 0.25 &  & CHZ & AVAX & 0.28 \\
    12 & ETH & AVAX & 0.60 &  & SOL & AVAX & 0.25 &  & SOL & AVAX & 0.28 \\
    13 & ETH & DOT & 0.59 &  & CHZ & BTC & 0.25 &  & LTC & BTC & 0.28 \\
    14 & ETH & BNB & 0.59 &  & ETH & AVAX & 0.24 &  & BTC & BCH & 0.28 \\
    15 & BTC & BNB & 0.58 &  & XRP & ETH & 0.24 &  & CHZ & BTC & 0.28 \\
    16 & SOL & LINK & 0.58 &  & XRP & CHZ & 0.23 &  & XRP & AVAX & 0.27 \\
    17 & LINK & DOT & 0.58 &  & LTC & BTC & 0.23 &  & ETH & DOGE & 0.27 \\
    18 & LINK & BTC & 0.56 &  & MYC & ENS & 0.23 &  & SOL & ETH & 0.26 \\
    19 & LINK & AVAX & 0.56 &  & DOT & BTC & 0.23 &  & SHIB & DOGE & 0.26 \\
    20 & LINK & ETH & 0.55 &  & XRP & SOL & 0.23 &  & DOGE & CHZ & 0.26 \\
    \bottomrule
    \end{tabular}
    \caption{Top 20 correlations between returns of cryptocurrencies traded on FTX.}
    \label{tab:HighCorr}
\end{table}

The summary statistics presented in Table \ref{tab:StatsFTX} delineate the market conditions in the three periods.
In the pre-collapse phase, the cryptocurrencies exhibited relatively low volatility and modest positive mean returns, except for FTT which showed a minimum negative return of -3.8596, indicating early signs of distress specific to the FTX token. The collapse phase presents a stark contrast, with all cryptocurrencies experiencing increased volatility and negative mean returns, highlighted by FTT’s dramatic mean return of $-0.0277$ and an exceptionally high standard deviation of $1.5709$, which is around 22 times the volatility observed in the pre-collapse phase. This period clearly reflects the acute market stress and investor panic triggered by the unfolding crisis. The full period combines these diverse phases, showing generally subdued mean returns and heightened volatility, which underscores the long-term impact of the crisis on market dynamics.

\begin{table}
    \centering
    \scriptsize
    \begin{tabular}{lrrrrrrr}
    \toprule
     & \textbf{AVAX} & \textbf{BTC} & \textbf{DOT} & \textbf{ETH} & \textbf{FTT} & \textbf{SOL} & \textbf{Market} \\
    \midrule
    \multicolumn{8}{l}{Panel A: pre-collapse} \\
    \midrule
    mean & 0.0008 & 0.0004 & 0.0005 & 0.0009 & 0.0003 & 0.0006 & 0.0004 \\
    std & 0.1023 & 0.0528 & 0.0855 & 0.0845 & 0.0724 & 0.0986 & 0.2598 \\
    min & -1.2175 & -0.7287 & -1.0151 & -0.9198 & -3.8596 & -1.1457 & -44.9316 \\
    Q1 & -0.0445 & -0.0204 & -0.0326 & -0.0365 & -0.0261 & -0.0480 & 0.0000 \\
    median & 0.0000 & 0.0000 & 0.0000 & 0.0000 & 0.0000 & 0.0000 & 0.0000 \\
    Q3 & 0.0446 & 0.0205 & 0.0327 & 0.0369 & 0.0264 & 0.0483 & 0.0000 \\
    max & 2.4216 & 1.6502 & 2.4874 & 1.9938 & 1.3012 & 2.1189 & 57.1662 \\
    \toprule
    \multicolumn{8}{l}{Panel B: collapse} \\
    \midrule
    mean & -0.0040 & -0.0026 & -0.0022 & -0.0018 & -0.0277 & -0.0071 & -0.0052 \\
    std & 0.3326 & 0.1964 & 0.3081 & 0.3734 & 1.5709 & 0.7396 & 1.7276 \\
    min & -8.1488 & -2.7816 & -9.1398 & -6.8753 & -29.9622 & -12.7876 & -172.0257 \\
    Q1 & -0.0811 & -0.0389 & -0.0721 & -0.0669 & -0.2097 & -0.1470 & 0.0000 \\
    median & 0.0000 & 0.0000 & 0.0000 & 0.0000 & 0.0000 & 0.0000 & 0.0000 \\
    Q3 & 0.0774 & 0.0353 & 0.0707 & 0.0599 & 0.1771 & 0.1331 & 0.0000 \\
    max & 6.3315 & 4.5145 & 5.4806 & 10.3006 & 32.3448 & 9.8197 & 299.0720 \\
    \toprule
    \multicolumn{8}{l}{Panel C: full period} \\
    \midrule
    mean & -0.0004 & -0.0003 & -0.0001 & 0.0002 & -0.0063 & -0.0013 & -0.0009 \\
    std & 0.1851 & 0.1063 & 0.1677 & 0.1964 & 0.7684 & 0.3706 & 0.8721 \\
    min & -8.1488 & -2.7816 & -9.1398 & -6.8753 & -29.9622 & -12.7876 & -172.0257 \\
    Q1 & -0.0494 & -0.0239 & -0.0344 & -0.0386 & -0.0359 & -0.0566 & 0.0000 \\
    median & 0.0000 & 0.0000 & 0.0000 & 0.0000 & 0.0000 & 0.0000 & 0.0000 \\
    Q3 & 0.0490 & 0.0235 & 0.0348 & 0.0383 & 0.0347 & 0.0560 & 0.0000 \\
    max & 6.3315 & 4.5145 & 5.4806 & 10.3006 & 32.3448 & 9.8197 & 299.0720 \\
    \bottomrule
    \end{tabular}
    \caption{Summary statistics of percentage log-returns of the 6 main cryptocurrencies traded on FTX.}
    \label{tab:StatsFTX}
\end{table}

\subsection{The network structure surrounding the collapse of the FTT node}
\label{sec:results1}

\begin{figure}
    \centering
    \includegraphics[width=.7\textwidth]{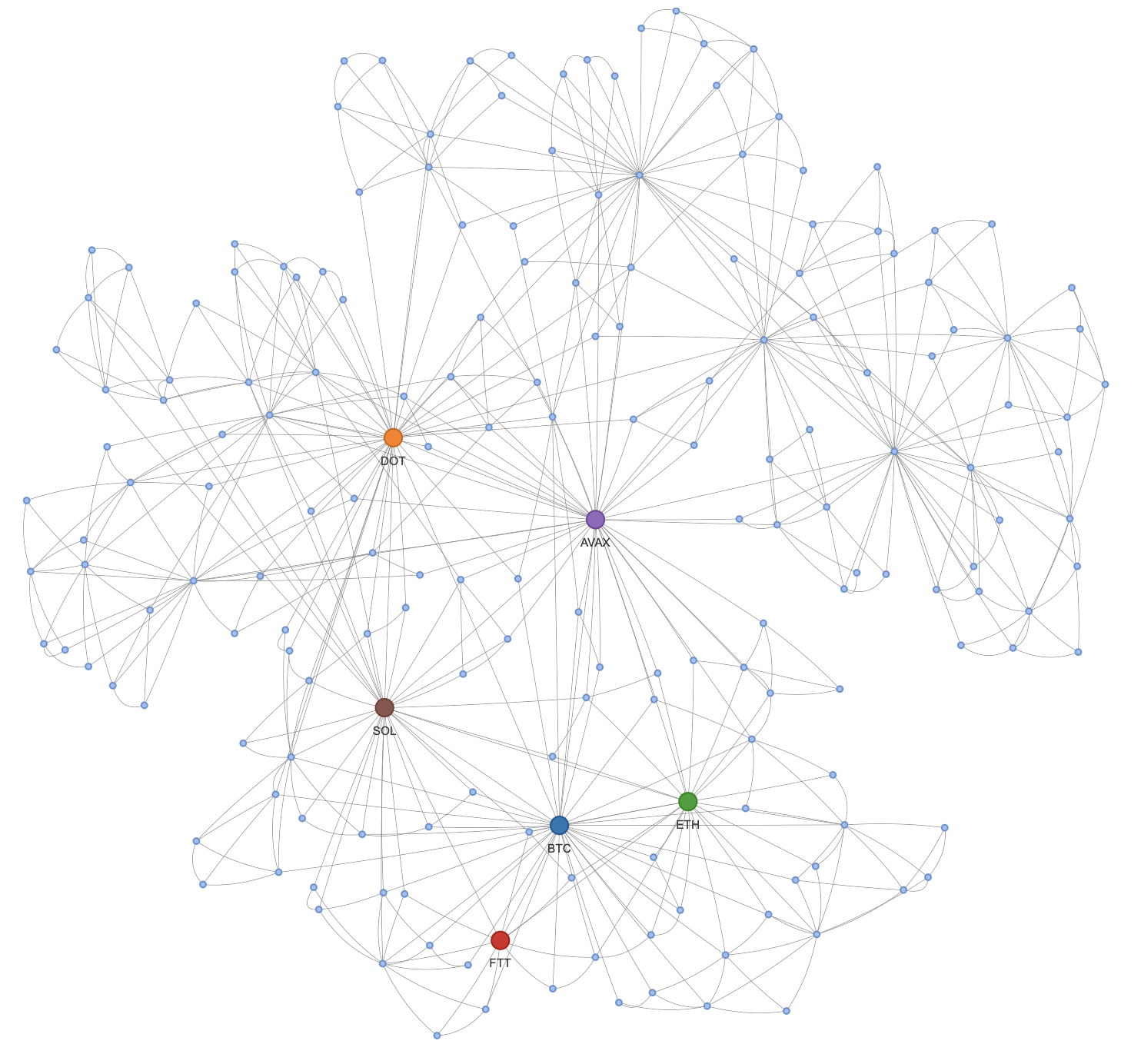}
    \includegraphics[width=.7\textwidth]{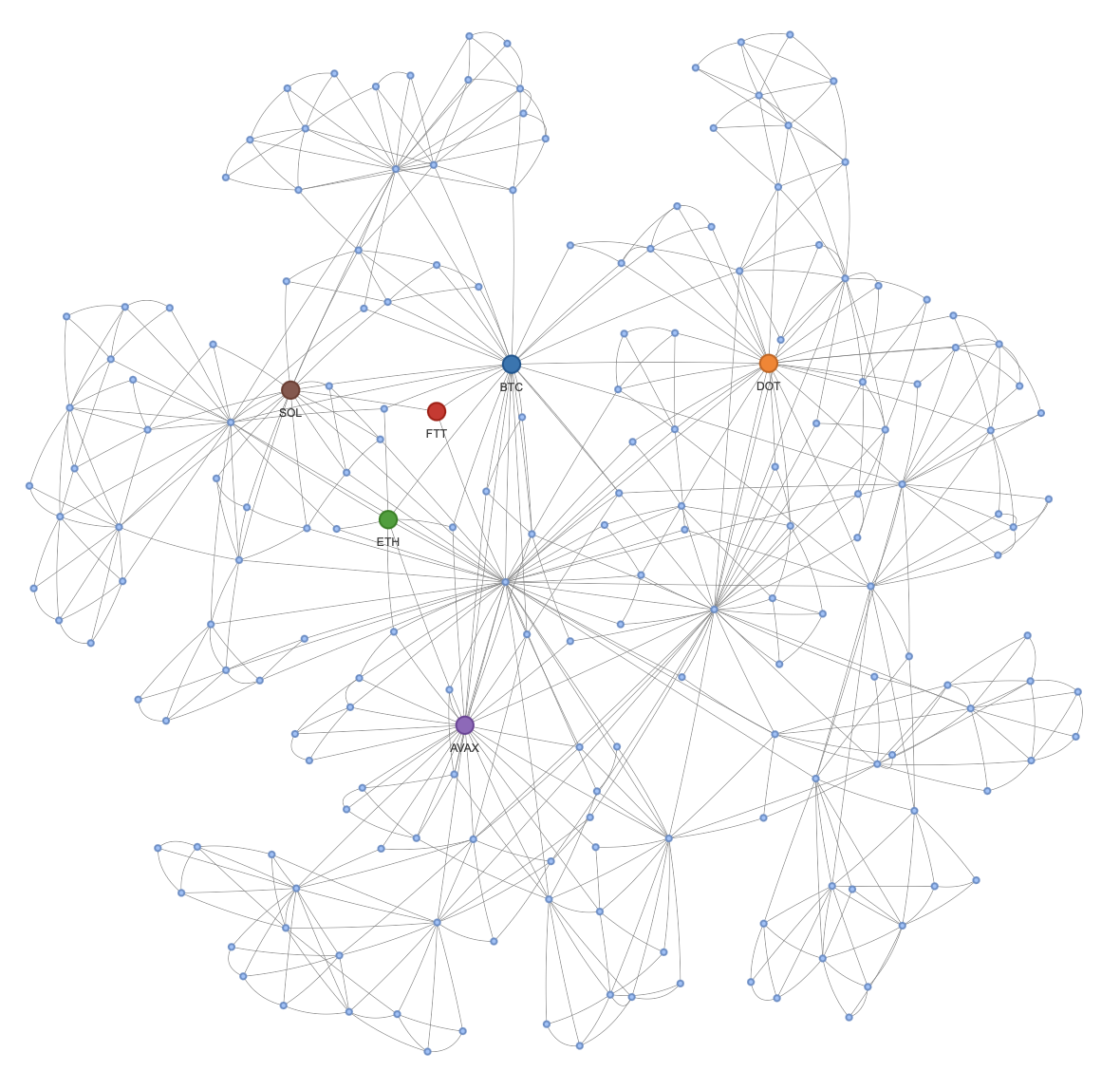}
    \caption{Maximal Filtered Planar Graph of the 195 cryptocurrencies from FTX market in the periods pre-collapse (top) and collapse (bottom). Colored nodes highlight the cryptocurrencies analysed.}
    \label{fig:network}
\end{figure}

To highlight changes in topological properties in the FTX market due to FTT's collapse, we construct two networks: one preceding the FTT's collapse and one during the collapse period after that event. 
Figure \ref{fig:network} shows the filtered 
networks of the FTX market in both periods pre and post event. 
We highlighted with different colors the nodes corresponding to the major cryptocurrencies that resulted in the highest correlations. 
However, the main topological changes cannot be detected by a simple visual network inspection. The analysis of the centrality measures computed for the cryptocurrency network particularly focusing on the FTX platform, reveals significant insights into the structural dynamics and influence patterns across the sample period.
Analyzing the normalized centrality measures — degree, closeness, and information centrality — provides an understanding of the node's importance, both prior to and following the market collapse. 

Table \ref{tab:centrality} presents the ranking of centrality measures 
of the top 20 cryptocurrencies in the FTX market, divided into the pre-collapse (Panel A) and collapse (Panel B) periods. We first observe that the centrality measures are close to zero, due to the network sparsity. 
However, the centrality measures well reflect the results of Table \ref{tab:HighCorr} revealing the most informative cryptocurrencies in the FTX market analysis. In the period preceding the FTT collapse (Panel A), AVAX and DOT emerged prominently across the centrality measures. AVAX, the token of Avalanche,  is one of the most promising emerging proof-of-stake (PoS) blockchain projects and perhaps the one that is followed with the greatest attention by developers and cryptocurrency investment experts. It shows a notable number of direct connections (normalized degree score $0.20103$) compared to other cryptocurrencies, suggesting a pivotal role in the network’s structure. Similarly, its closeness score ($0.00102$) underscores its centrality within the FTX network, being the most followed among cryptocurrencies as per its shortest distances from all the other nodes. However, in terms of information centrality, DOT takes precedence with a score of $0.06773$, reflecting its critical role in sustaining network efficiency in terms of rapid communication between nodes despite potential disruptions. Arguably, DOT is the native token of the Polkadot blockchain, which is considered among the most ambitious and innovative projects in the cryptocurrency arena, aiming to establish a decentralized web that enables various blockchains to interact and collaborate seamlessly.

BTC, consistently recognized for its market dominance, maintained notable centrality, reflective of its enduring influence and robust integration within the network. Interestingly, the presence of lesser-known cryptocurrencies like NEAR, a PoS protocol based on the concept of sharding, and SOL, a competitor of the Polkadot and Ethereum blockchains, in the top ranks across different measures highlights their emerging significance within the network’s architecture. ETH, the second-largest cryptocurrency in terms of market capitalization and liquidity, follows the rank among the most central nodes of the FTX network in terms of degree, closeness, and information centrality. Apart from NEAR, the centrality measures in the pre-collapse period show consistent results to the correlations ranking of Table \ref{tab:HighCorr} in Section \ref{sec:descriptive}. 

\begin{table}
    \centering
    \begin{adjustbox}{max width=\textwidth}
    \scriptsize
    \begin{tabular}{llrllrllr}
    \toprule
     & \multicolumn{8}{c}{Centrality Measures} \\\cmidrule{2-9}
     & \multicolumn{2}{c}{\textbf{Degree}} &  & \multicolumn{2}{c}{\textbf{Closeness}} &  & \multicolumn{2}{c}{\textbf{Information Centrality}} \\
    \midrule
    \multicolumn{9}{l}{Panel A: pre-collapse} \\
    \midrule
    1 & AVAX & 0.20103 &  & AVAX & 0.00102 &  & DOT & 0.06773 \\
    2 & DOT & 0.19588 &  & DOT & 0.00101 &  & AVAX & 0.06511 \\
    3 & BTC & 0.18041 &  & BTC & 0.00100 &  & BTC & 0.04789 \\
    4 & NEAR & 0.15979 &  & SOL & 0.00100 &  & SOL & 0.03962 \\
    5 & SOL & 0.15979 &  & ETH & 0.00099 &  & NEAR & 0.03362 \\
    6 & OMG & 0.14433 &  & NEAR & 0.00099 &  & ETH & 0.03283 \\
    7 & KNC & 0.12371 &  & BCH & 0.00098 &  & RAY & 0.02847 \\
    8 & ETH & 0.11856 &  & KNC & 0.00098 &  & TRX & 0.02816 \\
    9 & RAY & 0.09794 &  & AAVE & 0.00098 &  & OMG & 0.02761 \\
    10 & 1INCH & 0.08763 &  & LINK & 0.00098 &  & AAVE & 0.02690 \\
    11 & CHR & 0.07732 &  & 1INCH & 0.00097 &  & 1INCH & 0.02665 \\
    12 & TOMO & 0.07216 &  & RAY & 0.00097 &  & KNC & 0.02566 \\
    13 & AAVE & 0.06701 &  & BNB & 0.00097 &  & LINK & 0.02549 \\
    14 & LINK & 0.06186 &  & MATIC & 0.00097 &  & BCH & 0.02511 \\
    15 & \textbf{FTT} & 0.05670 &  & SRM & 0.00096 &  & SNX & 0.02372 \\
    16 & CRV & 0.05670 &  & \textbf{FTT} & 0.00096 &  & BNB & 0.02364 \\
    17 & TRX & 0.05670 &  & LTC & 0.00096 &  & \textbf{FTT} & 0.02355 \\
    18 & CRO & 0.05670 &  & UNI & 0.00096 &  & MATIC & 0.02300 \\
    19 & STETH & 0.05670 &  & ATOM & 0.00096 &  & SRM & 0.02280 \\
    20 & STSOL & 0.05155 &  & OMG & 0.00096 &  & UNI & 0.02272 \\
    \midrule
    \multicolumn{9}{l}{Panel B: collapse} \\
    \midrule
    1 & DOGE & 0.21134 &  & DOGE & 0.00276 &  & DOGE & 0.05907 \\
    2 & CHZ & 0.17010 &  & BTC & 0.00274 &  & BTC & 0.05406 \\
    3 & BTC & 0.14433 &  & AVAX & 0.00266 &  & XRP & 0.04525 \\
    4 & AVAX & 0.14433 &  & CHZ & 0.00264 &  & AVAX & 0.04434 \\
    5 & DOT & 0.13402 &  & XRP & 0.00258 &  & DOT & 0.03845 \\
    6 & XRP & 0.10309 &  & DOT & 0.00258 &  & CHZ & 0.03704 \\
    7 & MATIC & 0.10309 &  & ETH & 0.00257 &  & STG & 0.03523 \\
    8 & SOL & 0.09278 &  & SOL & 0.00252 &  & MATIC & 0.03458 \\
    9 & ATOM & 0.08247 &  & LTC & 0.00248 &  & ENS & 0.02852 \\
    10 & LTC & 0.08247 &  & RSR & 0.00241 &  & LDO & 0.02840 \\
    11 & GT & 0.07216 &  & BCH & 0.00241 &  & LTC & 0.02569 \\
    12 & STG & 0.07216 &  & LINK & 0.00240 &  & WFLOW & 0.02553 \\
    13 & LRC & 0.06701 &  & ATOM & 0.00239 &  & AXS & 0.02402 \\
    14 & CRV & 0.06701 &  & \textbf{FTT} & 0.00237 &  & ATOM & 0.02291 \\
    15 & AAVE & 0.06186 &  & MATIC & 0.00235 &  & BAND & 0.02277 \\
    16 & ALICE & 0.05670 &  & UNI & 0.00226 &  & IP3 & 0.02157 \\
    17 & WFLOW & 0.05670 &  & CRV & 0.00224 &  & SOL & 0.02157 \\
    18 & BAND & 0.05670 &  & USDT & 0.00224 &  & C98 & 0.02126 \\
    19 & SPELL & 0.05155 &  & BAND & 0.00223 &  & ETH & 0.02076 \\
    20 & IP3 & 0.05155 &  & APT & 0.00223 &  & OMG & 0.02061 \\
    \bottomrule
    \end{tabular}
    \end{adjustbox}
    \caption{Ranking of the top 20 cryptocurrency in terms of degree, closeness, and information centrality measures in the FTX market}
    \label{tab:centrality}
\end{table}

The collapse phase delineated a stark transformation in network structure 
(see Figure \ref{fig:network}), with Dogecoin (DOGE), one of the first altcoins\footnote{i.e., alternative coins to Bitcoin.} created as a fast and instant payment system based on the Litecoin blockchain architecture, leading in all centrality measures. Its degree centrality (0.21134) notably increased, paired with the highest closeness (0.00276) and information centrality (0.05907) scores, indicating a surge in its network influence post-collapse. This shift might suggest a realignment of network structures where market participants pivot towards alternative nodes during periods of instability. Ripple (XRP) and Litecoin (LTC), absent in the centrality ranking of the pre-collapse period, also gained more importance in terms of network centrality and information dissemination during the collapse period, a signal of investors’ reliance on well-established cryptocurrencies. Indeed, BTC maintained its predominant position, albeit with reduced scores compared to DOGE, signaling its resilience and foundational role within the network even amidst market upheavals. Other cryptocurrencies like AVAX and DOT also remained influential, though with adjusted rankings, reflecting a reconfiguration of inter-node relations and dependencies. 

Before the collapse of the FTX market, FTT was solidly positioned within the top 20 in degree centrality, signifying a robust number of direct connections, i.e., strong correlations within a high number of cryptocurrencies, and an important role within the network’s structure. This initial placement highlighted its prominence and influence among investors and traders on the platform. However, after the collapse of FTX market, we observe that FTT experienced a significant reduction in its degree centrality, indicating a loss of its direct connections and, therefore, high correlations with other cryptocurrencies within the network. 
Another interesting result emerges from the ranking of the closeness.
Indeed, results in Table \ref{tab:centrality} show that the collapse of FTT has led to an increase in closeness centralities as a consequence of the average decrease of the geodesic distances between the nodes of the network. This shift suggests a decrease in its active integration and participation within the trading environment, likely due to diminished investor confidence and a restructuring of the network’s dynamics. Despite this decline in degree and information centralities, FTT is still present in the closeness centrality's ranking. This result derives from the fact that FTT keeps few connections in the network and its neighbors are central in terms of degree, closeness and information centrality.\footnote{The FTT's neighbors after its collapse are: BTC, SOL, and DOGE} This could be economically interpreted as a sign of investors’ need to readjust their positions during times of crisis, reflecting a scenario of ongoing adjustments and liquidations by investors trying to navigate the market’s volatility. Interestingly, the information centrality for FTT post-collapse suggests an absence of significant information flow through the FTX's token, underscoring a scenario where investors were likely in a rush to disassociate from the token amid its collapse. This lack of flowing information indicates that while FTT retained some network closeness—perhaps due to its previous importance and lingering transactions—the effective communication and utility of FTT within the network had substantially diminished, mirroring the broader crisis impacts on its value and operational significance.

\subsection{Pattern of information flow surrounding the collapse of the FTT node}
\label{sec:results2}

\begin{figure}
    \centering
    \includegraphics[width=\textwidth]{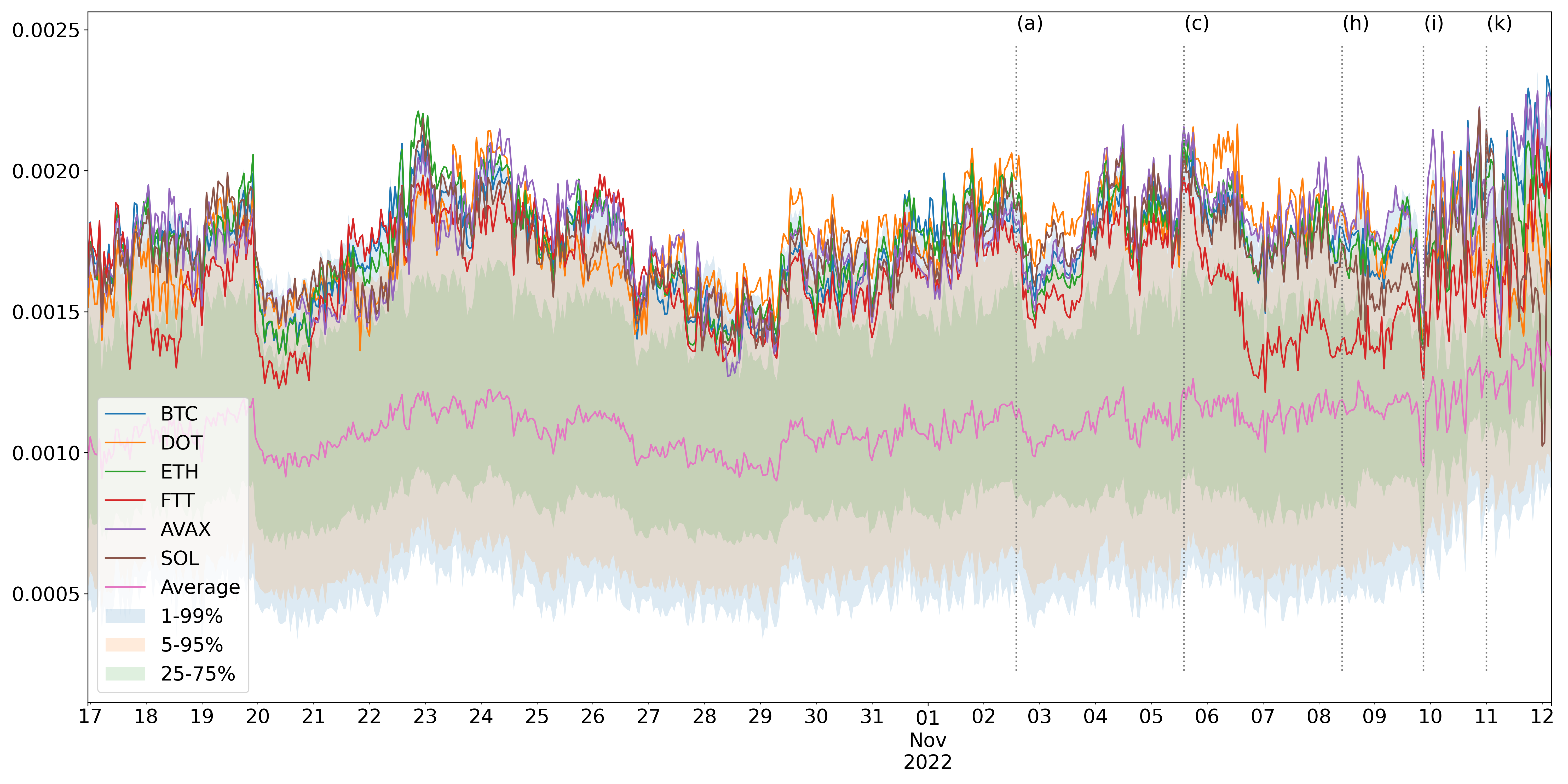}
    \includegraphics[width=\textwidth]{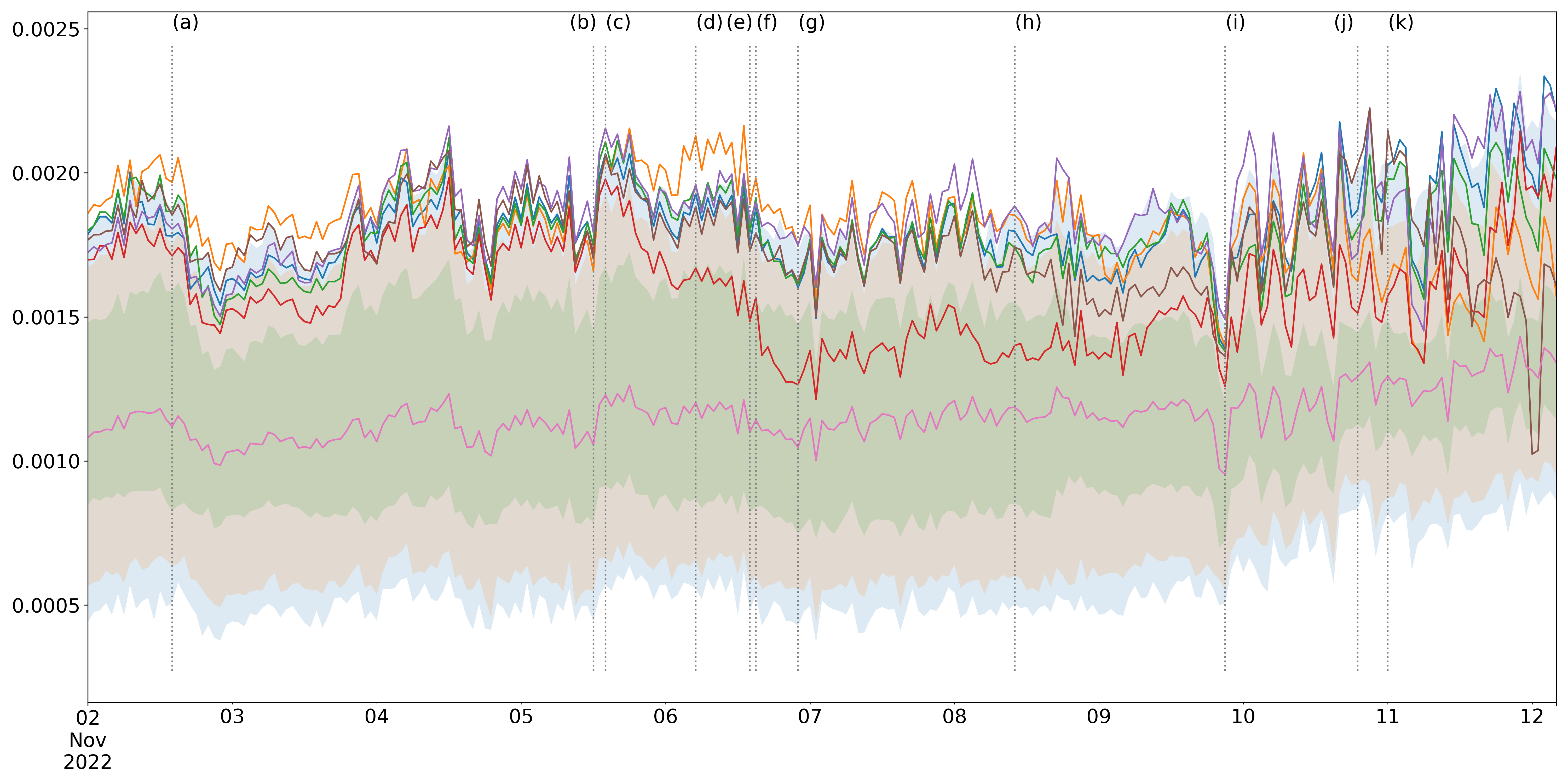}
    \caption{Rolling closeness centrality on FTX network in the entire sample period (top) and the zoomed-in collapse period (bottom).}
    \label{fig:rolling_closeness}
\end{figure}

To deepen the analysis of the previous Sections, we capture the evolving patterns in the FTX network by means of rolling windows. In particular, we construct daily rolling window networks as in Sect.\ \ref{sec:network_construction} with 24-hour windows at intervals of one hour. Based on the results of the previous Sections, we focus on the closeness and the information centrality as the most informative centrality measures. We plot in Figure \ref{fig:rolling_closeness}  the daily rolling closeness centrality measure\footnote{The normalized version is represented by Figure \ref{fig:rolling_closenessN} in \ref{appendix_figures})} over time for the major cryptocurrencies of the FTX network. This measure is crucial for understanding how well connected (in terms of reachability) a node is, not only through direct links but with respect to the entire network’s structure. Figure \ref{fig:rolling_closeness} 
illustrates that Bitcoin (BTC) consistently maintained the highest score, indicative of its central role in the network.

The prominence of BTC throughout the period illustrates its foundational status within the cryptocurrency network, maintaining efficient information dissemination capabilities despite market upheavals. Overall, the average closeness centrality (pink line) of the FTX network saw a slight increase after the announcement of Binance dropping out from the acquisition deal (Event (i)). This signals a reduction of the distance between pairs of nodes and therefore, an improvement in the reachability\footnote{In graph theory, the reachability refers to the ability of a node $i$ to reach all the other nodes in the network.} between cryptocurrencies after the disruption of the FTT. Each remaining colors correspond to the other main cryptocurrencies analyzed, while the three ranges comprise the percentiles as described in the Figure legend.

As previously discussed, specific events had pronounced impacts on the centrality measures of other cryptocurrencies, especially FTT. 
The revelation of compromised balance sheet of Alameda Research, containing significant amounts of FTT, precipitated an immediate reaction: it can be easily noticed by the drop in FTT’s centrality on November $2^{nd}$. This initial impact was compounded by subsequent market movements and public statements from key market players. The series of events from November $5^{th}$ through $6^{th}$, involving the large-scale movement of FTT by Binance and Binance’s CEO’s public warnings, sparked notable fluctuations in FTT’s centrality. This is even more evident in Figure \ref{fig:rolling_closenessN} of \ref{appendix_figures} when normalizing by the average centrality of the entire FTX network. These changes underscored the market’s sensitivity to news and actions by major stakeholders, reflecting real-time adjustments within the network’s structure. 

The heightening uncertainty following conflicting public statements about the insolvency of FTX and Alameda Research and the dramatic fallout from the failed Binance acquisition deal further influenced the network dynamics. Closeness centrality values for all the cryptocurrencies analyzed, particularly after Event (i) in Table \ref{tab:events}, displayed erratic movements that mirrored the tumultuous information environment. The operational disruptions at FTX, highlighted by the halting and partial reopening of withdrawals, followed by the collapse of the FTT token, were critical as well. These events led to a pronounced and sustained decline in FTT’s closeness centrality, but also to a major centrality volatility in other major cryptocurrencies within the network. This decline was stark against the backdrop of FTX filing for bankruptcy and closing operations, where FTT’s centrality reached its nadir by November $11^{th}$, 
showing its reduced role in the network.

\begin{figure}
    \centering
    \includegraphics[width=\textwidth]{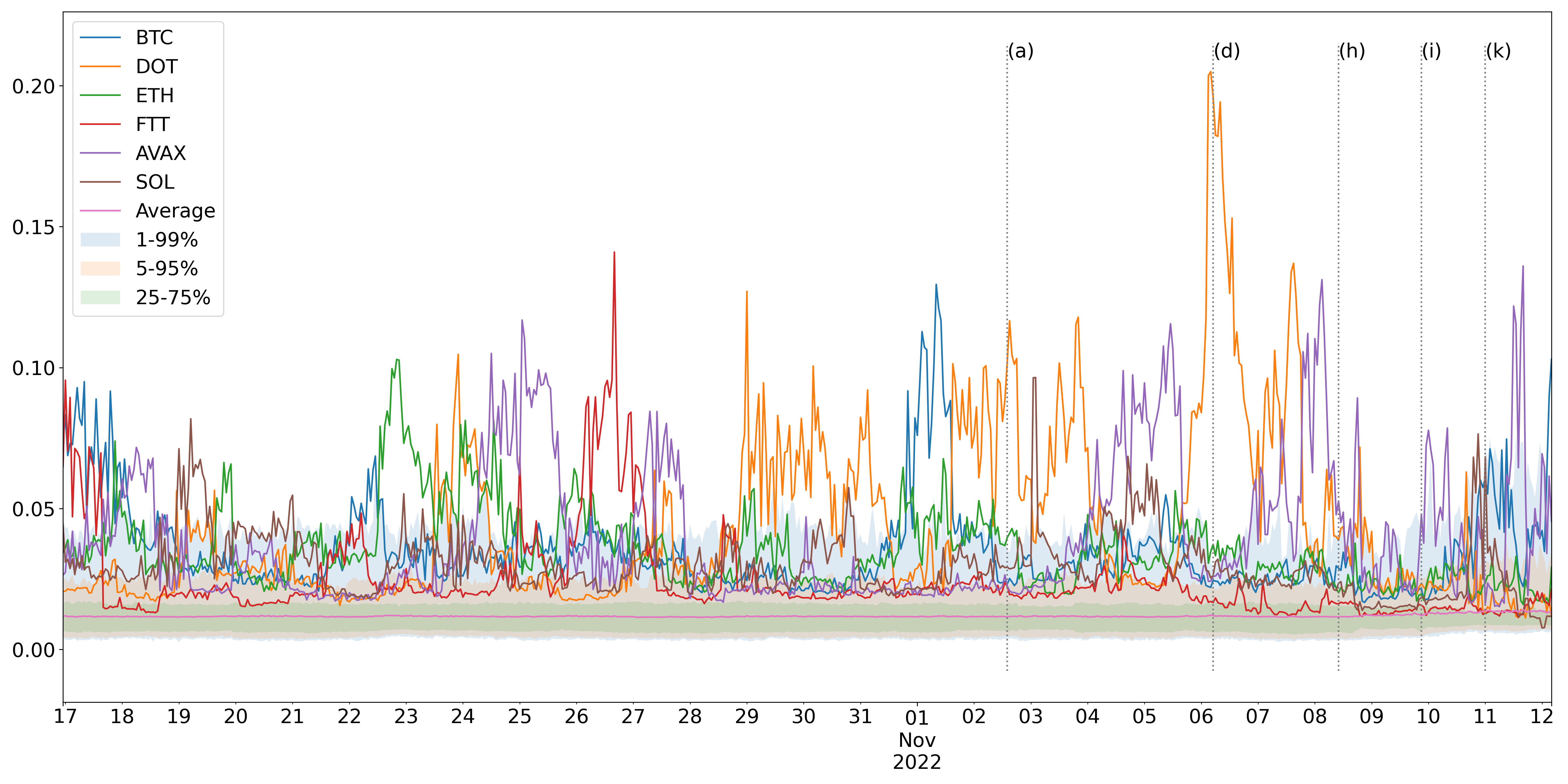}
    \includegraphics[width=\textwidth]{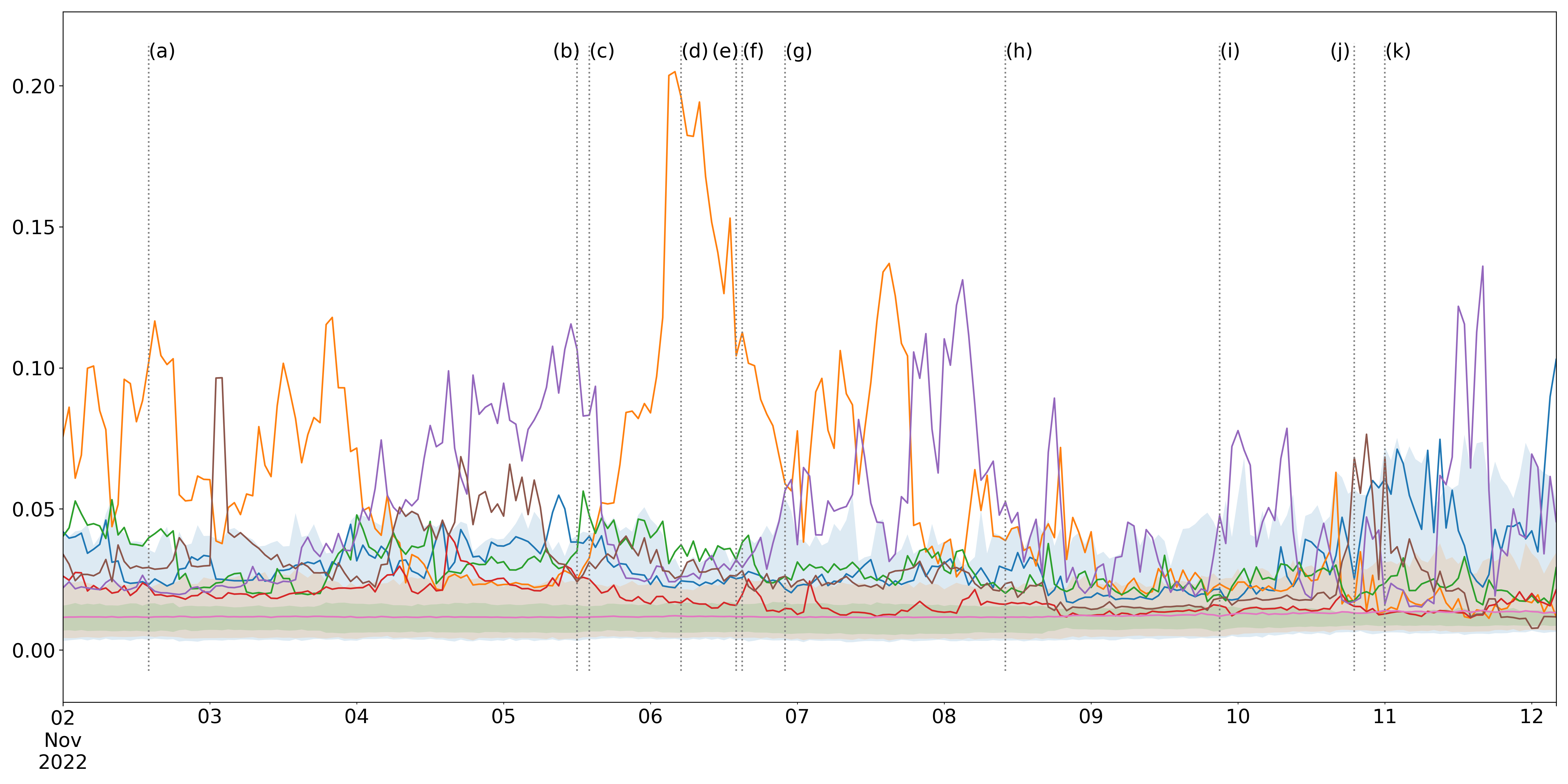}
    \caption{Rolling information centrality on FTX network in the entire sample period (top) and the zoomed-in collapse period (bottom).}
    \label{fig:rolling_information}
\end{figure}

Figure \ref{fig:rolling_information} shows, instead, the patterns of nodes' information centrality measure. Information centrality, a measure of the efficiency of information flow through the shortest paths in the network, showcases distinct patterns in response to the events leading to the collapse of FTT. The plot shows spikes in DOT’s information centrality around the time when concerns about Alameda Research’s balance sheet surfaced (Events (a) and (d) in Table \ref{tab:events}). This highlights the importance of DOT within the network as a safe-haven vehicle through which to disseminate information. The same happens to AVAX, which saw major spikes throughout the FTT collapse period as well. On the other hand, this could also indicate temporary scrutiny of FTT within the network. The reason can likely be due to increased transactions and information flow as market participants assessed the implications of the news. Perhaps cryptocurrencies like DOT and AVAX are likely traded less for their complexities than more popular ones, such as BTC and ETH, but they still represent central points for network structure, suggesting the importance of more stable protocols over crypto-assets popularity driven by less sophisticated investments.

\begin{figure}
    \centering
    \includegraphics[width=\textwidth]{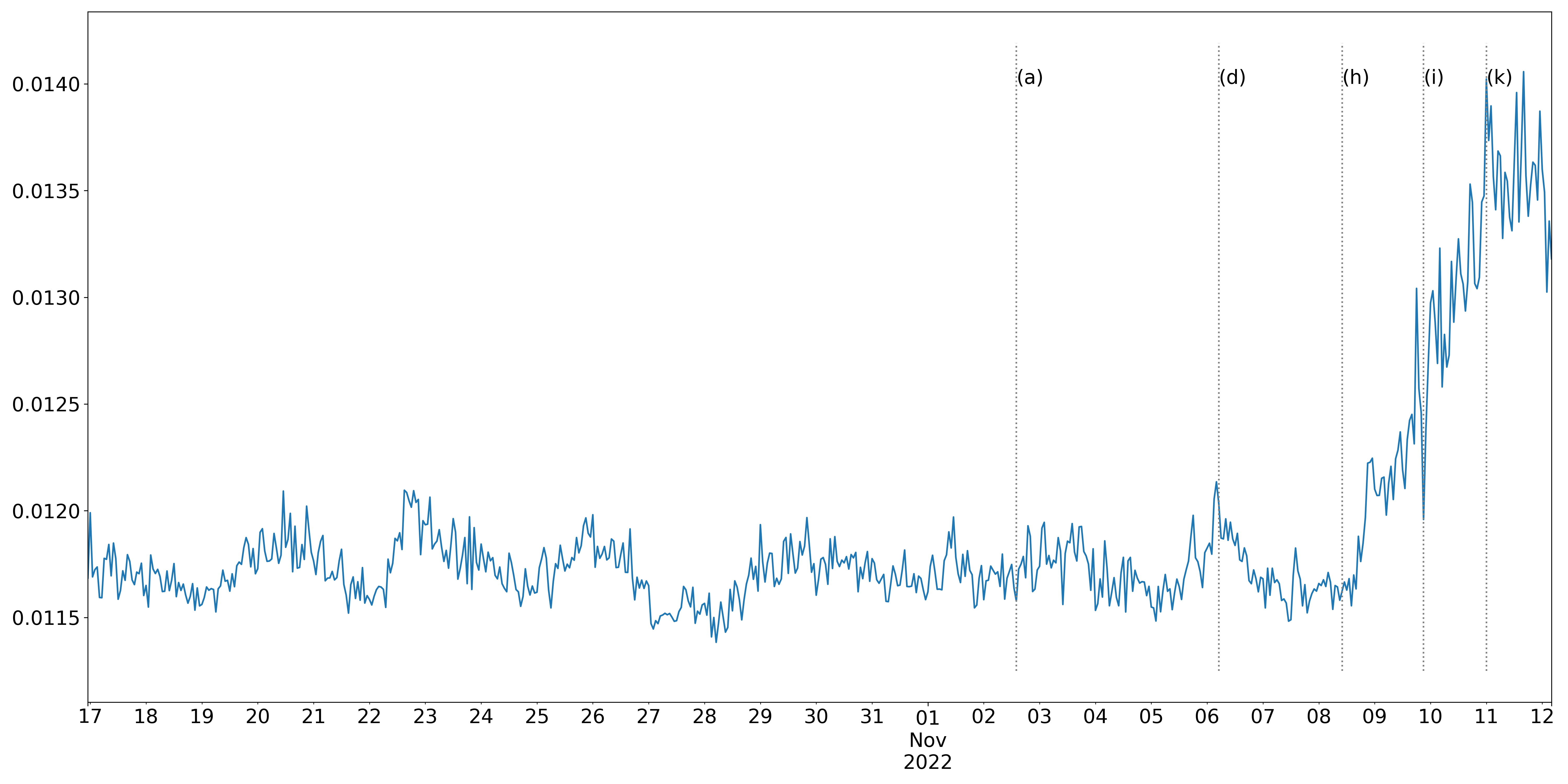}
    \includegraphics[width=\textwidth]{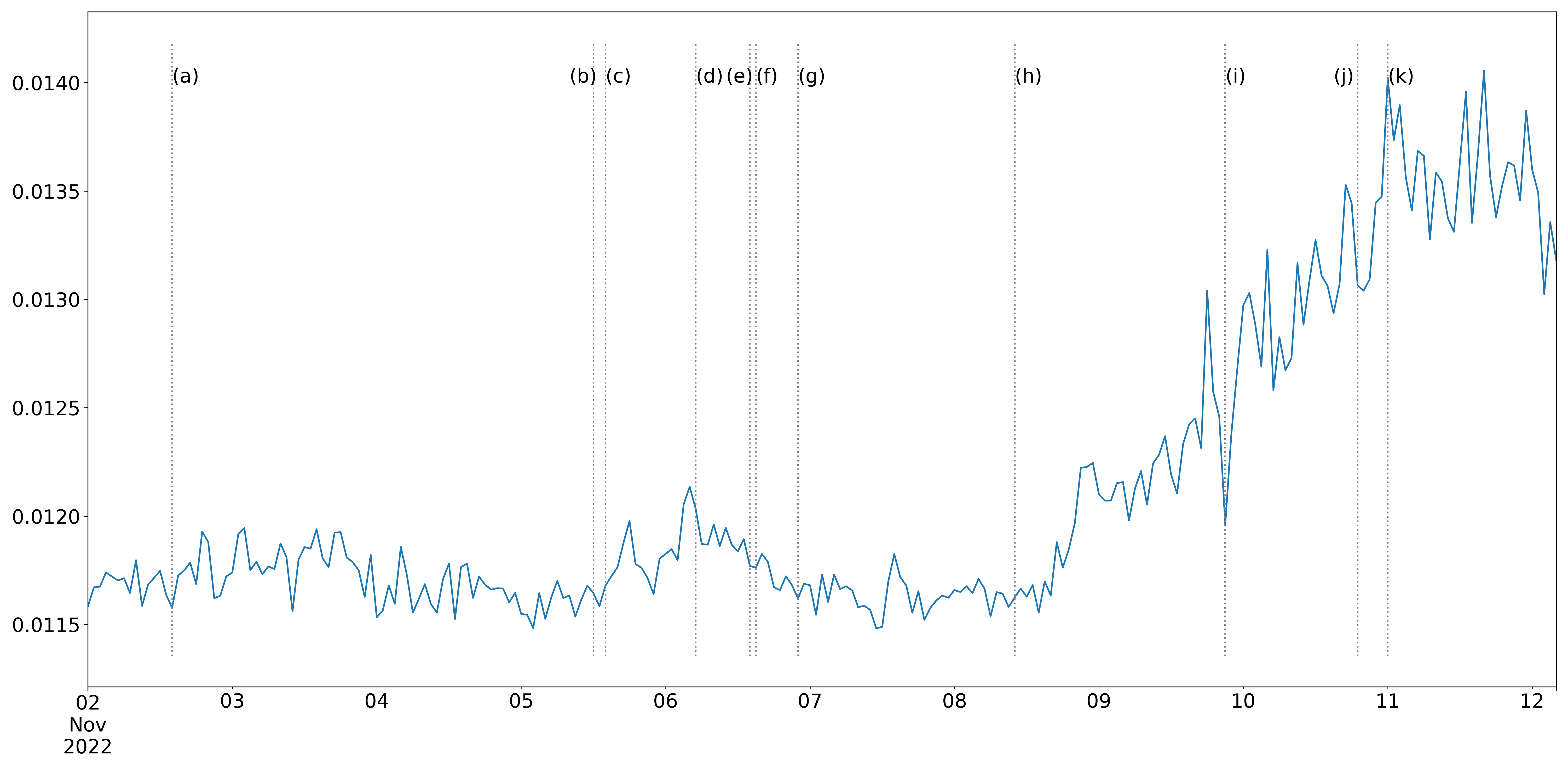}
    \caption{Average rolling information centrality on FTX network in the entire sample period (top) and the zoomed-in collapse period (bottom).}
    \label{fig:rolling_average_inf}
\end{figure}

Throughout this turbulent period, other cryptocurrencies such as BTC, ETH, and SOL also experienced shifts in their information centrality measures, though less dramatically than FTT. This pattern underscores the broader impact of the crisis, affecting the entire cryptocurrency market as participants adjusted their strategies in response to the unfolding events. Finally, in Figure \ref{fig:rolling_average_inf} we plot the average information centrality measure of the entire FTX network $c_I(G)$, showing the overall trend of the information flow during the sample period. As it is readily apparent, there is a structural break after the Event (h) in Table \ref{tab:events}, showing that the impossibility of withdrawing digital assets from the exchange led investors to cautiously divest and re-allocate their positions in the face of the crisis, a result consistent with the findings in \cite{galati_market_2024}. While the weeks before the first announcement and the period during the collapse of FTX’s native token show a stable flow of information within the network, there was substantial informed trading after the exchange halted funds withdrawals to avoid a “bank run”.

The observed centrality trends from the figures provide detailed insights into how information is disseminated and processed within the cryptocurrency network in response to crisis events. These dynamics are indicative of the market’s reactive nature and the critical role of information centrality in understanding the mechanisms of crisis management and response in decentralized financial systems. It is interesting to see how well the centrality measures used in this study reflect the market behaviors surrounding each of the events considered, signaling the appropriateness of the methodology employed for the proposed investigation.

\subsection{The market responses in the Binance network}
\label{sec:results3}

Consistent with the market comparison in \cite{galati_market_2024} and the dominance results of \cite{vidal-tomas_ftxs_2023}, we lastly explore the centrality measures across major cryptocurrencies within the Binance network during the FTX crisis. In particular, similarly to the previous section, we construct daily rolling window networks at intervals of one hour. Each network has $n=324$ nodes and it is constructed as in Sect.\ \ref{sec:network_construction}. Our aim is to provide a better overall understanding of the decentralized financial system dynamics that underpin the distribution and reception of information in times of financial upheaval. Given Binance’s significant market presence, this analysis serves as a proxy for broader market behaviors and offers insights into how cryptocurrencies interact and respond within one of the largest trading environments.

\begin{figure}
    \centering
    \includegraphics[width=\textwidth]{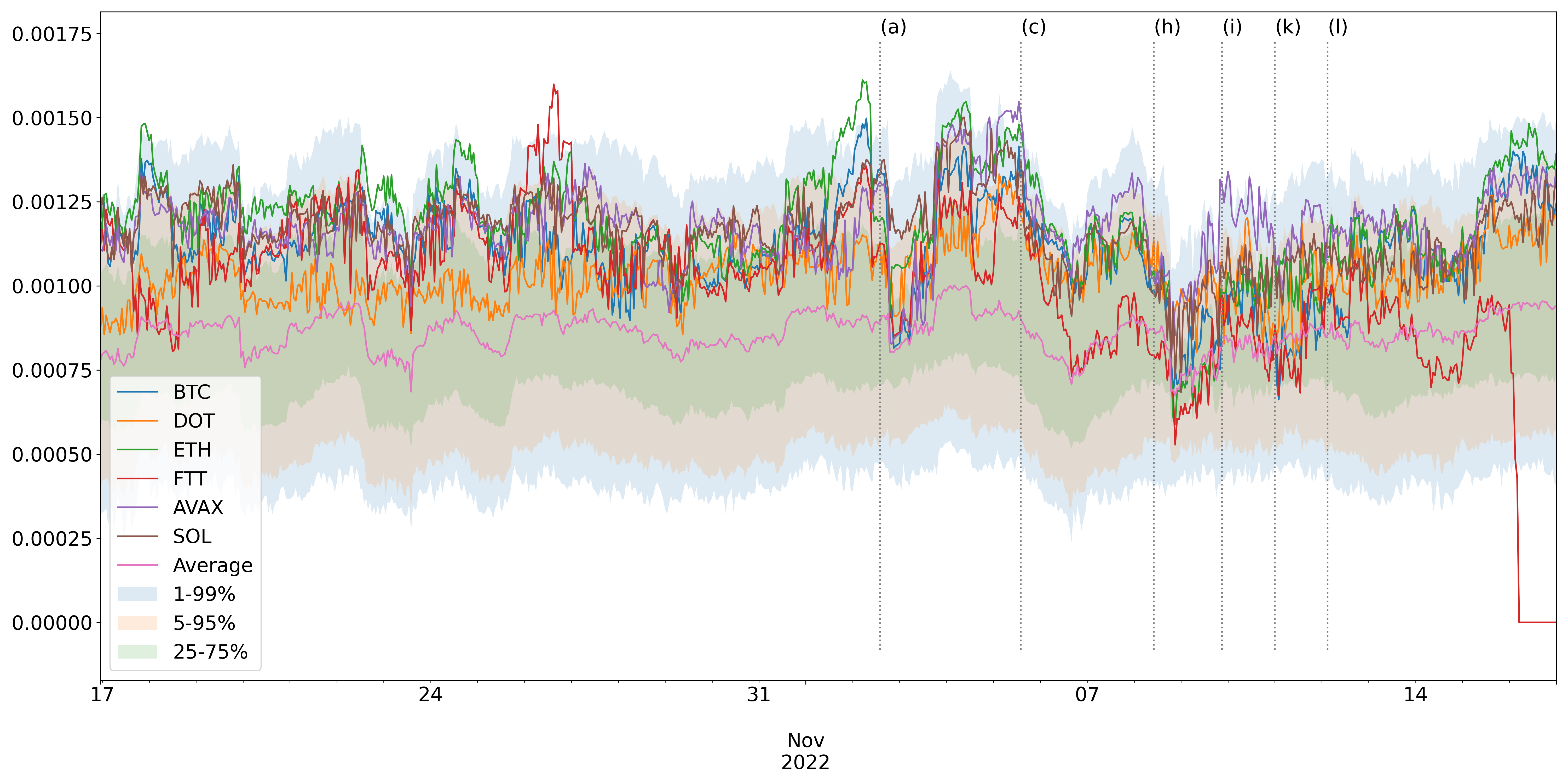}
    \includegraphics[width=\textwidth]{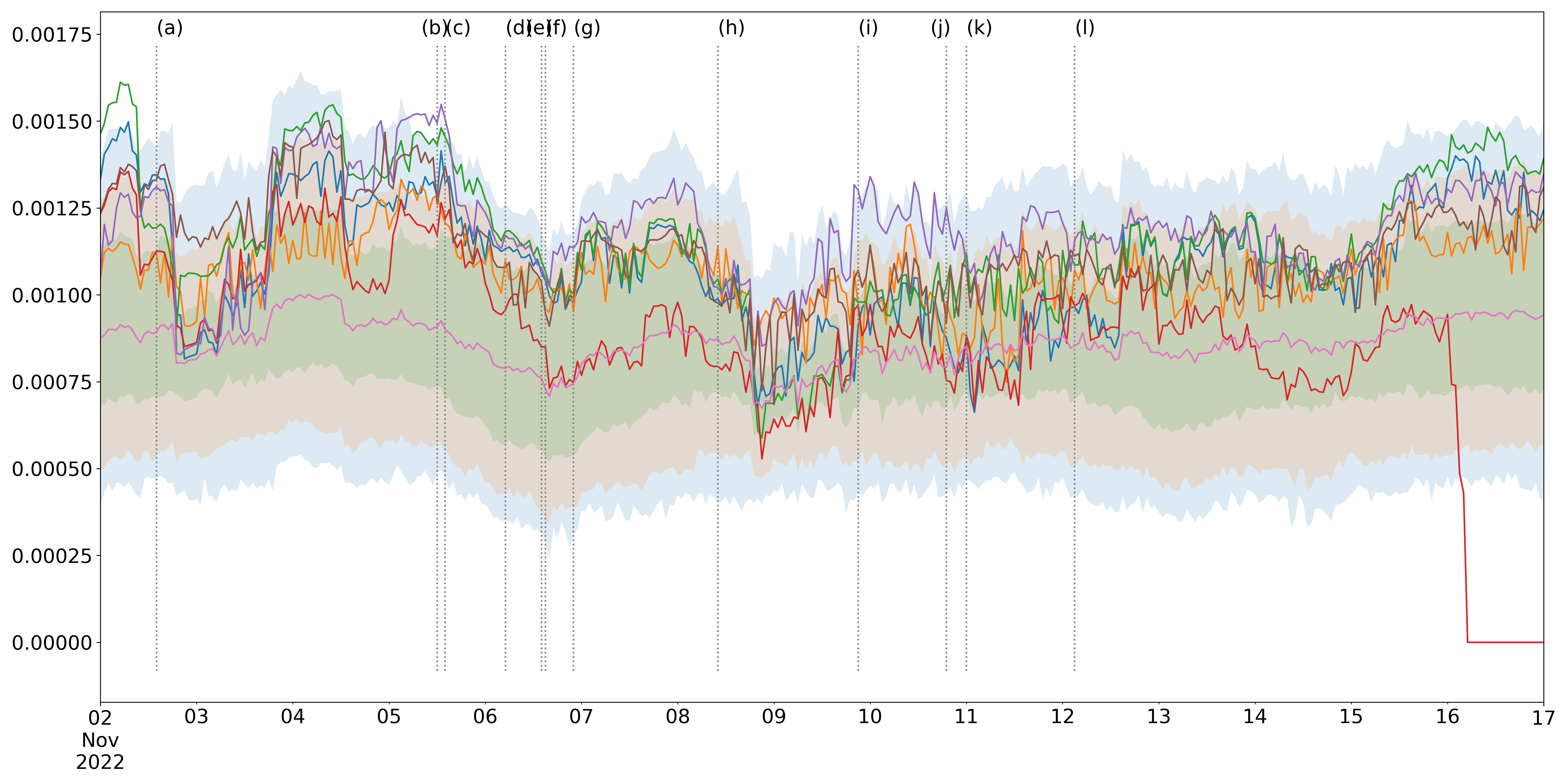}
    \caption{Rolling closeness centrality on Binance network in the entire sample period (top) and the zoomed-in collapse period (bottom).}
    \label{fig:rolling_closenessB}
\end{figure}

Figure \ref{fig:rolling_closenessB} (as well as its normalized version in Figure \ref{fig:rolling_closenessNB} in \ref{appendix_figures}) clearly depicts the shifts in closeness centrality that occurred in response to the pivotal events mentioned above. Throughout the observed period, BTC demonstrates relatively high and stable closeness centrality, indicative of its central role in the cryptocurrency market. This stability reflects Bitcoin’s broad acceptance within the whole market as a central node through which information and transactions are disseminated. Besides, 
we notice the resilience of the overall Binance network shown after the Event (h), as all the closeness reversed and started to increase again. This is consistent with the findings of \cite{galati_market_2024} documenting an improvement of Binance market quality upon the halt of funds withdrawal on FTX. 

Particularly for FTT, the figures show a distinct pattern in closeness centrality, peaking around critical events before exhibiting a dramatic fall as the crisis escalated. This trend is even more visible on the Binance network, within which the FTT closeness is downward below the overall closeness average of the network. This is crucial as it reflects how such crises can influence related cryptocurrencies within a large trading platform like Binance, which is supposedly not connected with the bankrupted FTX exchange. Contrastingly, most popular cryptocurrencies, like BTC and ETH, displayed more stable closeness centrality during the crisis, suggesting a retained level of trust and stability despite the surrounding turmoil. 

While FTT shows significant centrality spikes (before the first announcement) and subsequent declines (during the collapse), other cryptocurrencies like AVAX and SOL display varying levels of centrality changes. These variations could be indicative of differing levels of exposure to systemic shocks. Additionally, DOT shows similar patterns albeit seemingly less volatile in the Binance network compared to FTX. Before the onset of the crisis, the closeness centrality of all tracked cryptocurrencies exhibits a somewhat synchronized pattern, suggesting a tightly interlinked market structure. As the crisis unfolds, the divergence in centrality measures becomes more pronounced, underscoring a shift in how these cryptocurrencies interact and influence each other within the network.

\begin{figure}
    \centering
    \includegraphics[width=\textwidth]{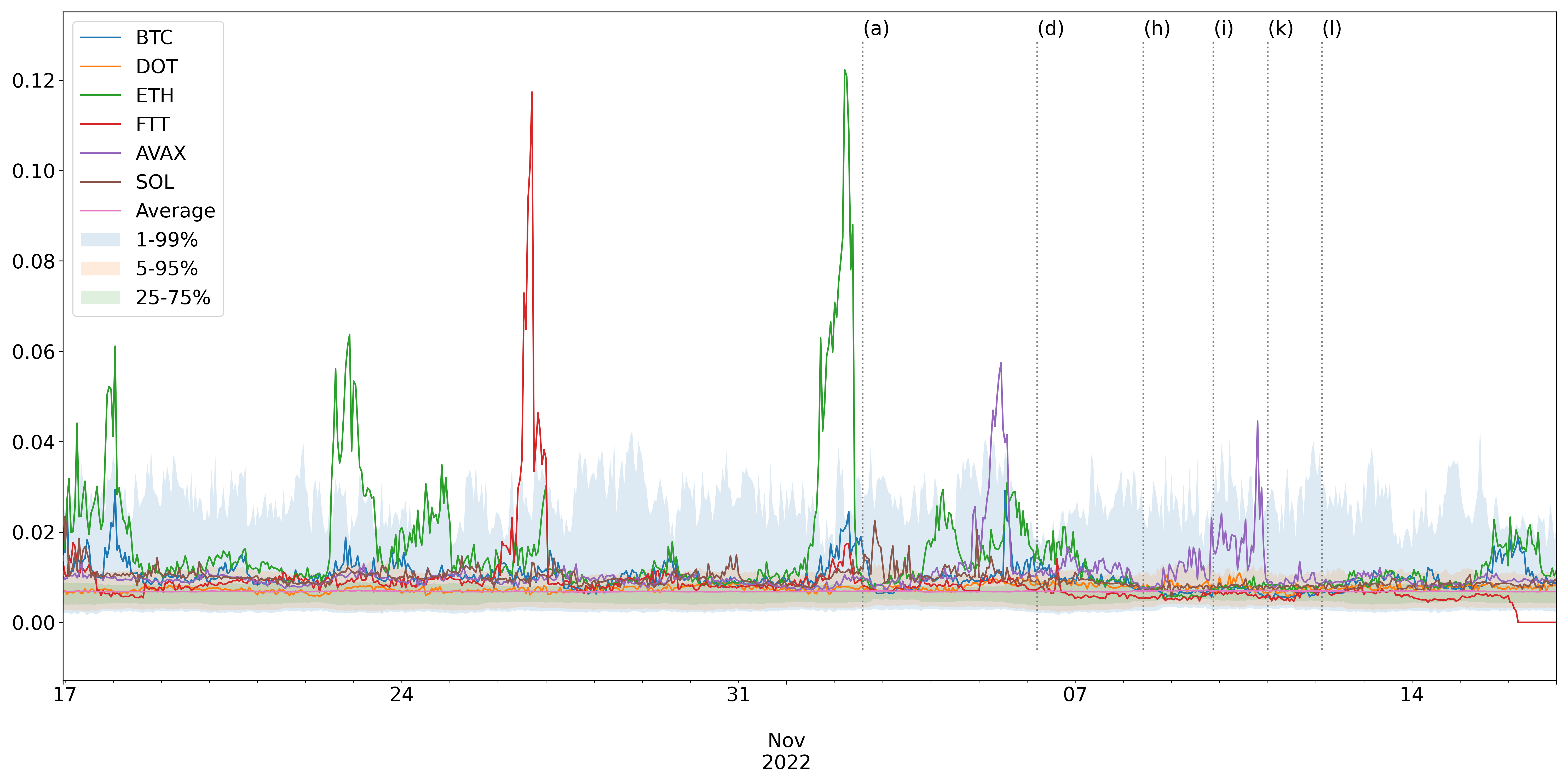}
    \caption{Rolling information centrality on Binance network.}
    \label{fig:rolling_informationB}
\end{figure}

Figure \ref{fig:rolling_informationB} reveals noticeable fluctuations in information centrality as the FTX crisis unfolded. Interestingly, as in Figure \ref{fig:rolling_information}, FTT exhibited sharp centrality peaks coinciding with a supposedly market event that anticipated the crisis. 
The same happens for ETH, showing what can arguably be described as insider trading, or more generally, an anticipation of information leading to the overall crisis.\footnote{ETH is arguably the most central node within the Binance network in terms of the measures considered.} Its spike, just before the first public announcement of FTX's crisis, 
might be seen as a signal of the information spreading through Binance to the overall market.

ETH and AVAX had the highest information centrality on Binance during the sample period, particularly after Event (a). Such spikes in centrality reflect their role in information spreading due to a surge in transaction activities associated with these tokens. Overall, the patterns of the other major cryptocurrencies do not show significant changes, highlighting their non-primary role in spreading information. 
This emphasizes the role of Binance’s network in mediating these dynamics even though the crisis incurred on another exchange.

\section{Conclusions}
\label{sec:conclusion}

This study examined the anatomy of information flow in the bankrupted FTX cryptocurrency exchange by underscoring the distinct role of centrality measures in capturing the critical dynamics within financial networks during periods of market instability. Through an exploration of degree, closeness, and information centrality, our findings reveal how the market structure and the efficacy of information dissemination within such networks evolve in response to substantial financial disruptions. In this study, we also elucidate the dynamic 
nature of centrality within the cryptocurrency markets more broadly, influenced by both internal network adjustments and external market conditions. 

Our analysis illustrates that path-based centrality measures in financial networks not only offer insights into the immediate impacts of such crises but also highlight the broader structural vulnerabilities and resilience within the cryptocurrency market. The rapid shifts in centrality metrics observed in response to the unfolding crisis at FTX reflect the market’s sensitivity to both internal network changes and external economic pressures. This study exposes the interconnected nature of cryptocurrency markets, where the failure of a single node such as FTT can lead to widespread repercussions across different networks, affecting investor behavior and market stability. Importantly, the resilience of certain cryptocurrencies like Bitcoin, which maintained high centrality throughout the crisis, contrasts sharply with the volatility in centrality measures of other tokens directly implicated in the crisis, like FTT. This disparity underscores the importance of robust network positions and the dangers of over-reliance on individual market players within decentralized financial systems.

Finally, the insights derived from this research have significant implications for stakeholders across the financial spectrum—from investors and analysts to regulators and policymakers. Understanding the centrality dynamics within such networks can aid in developing more resilient financial structures and inform strategic decisions aimed at mitigating risks associated with market centralization and the cascading effects of financial crises. This study, therefore, not only contributes to the academic debate on complex networks and financial markets but also provides practical frameworks for assessing and enhancing the stability of the cryptocurrency ecosystem.

\section*{Declaration of Competing Interest}
The authors declare that they have no known competing financial interests or personal relationships that could have appeared to influence the work reported in this paper.

\section*{Acknowledgement}
Luca Galati thanks Refinitiv, an LSEG business, which provided him with access to data and technical assistance. \\ 
Riccardo De Blasis is member of the Gruppo Nazionale Calcolo Scientifico-Istituto Nazionale di Alta Matematica (GNCS-INdAM). \\
Rosanna Grassi acknowledges financial support from the European Union – NextGenerationEU. Project PRIN 2022 “Networks: decomposition, clustering and community detection” code: 2022NAZ0365 - CUP H53D23002510006. Rosanna Grassi is a member of GNAMPA-INdAM. \\
Giorgio Rizzini gratefully acknowledges financial support from “SoBigData.it”, which receives funding from the European Union – NextGenerationEU – PNRR – Project: “SoBigData.it – Strengthening the
Italian RI for Social Mining and Big Data Analytics” – Prot. IR0000013 – Avviso n. 3264 del
28/12/2021.
Giorgio Rizzini acknowledges partial support by the European Program scheme “INFRAIA-01-2018-2019:
Research and Innovation action”, grant agreement n. 871042 “SoBigData++: European Integrated
Infrastructure for Social Mining and Big Data Analytics”.

\FloatBarrier
\appendix

\section{Additional figures}
\label{appendix_figures}

\begin{figure}
    \centering
    \includegraphics[width=\textwidth]{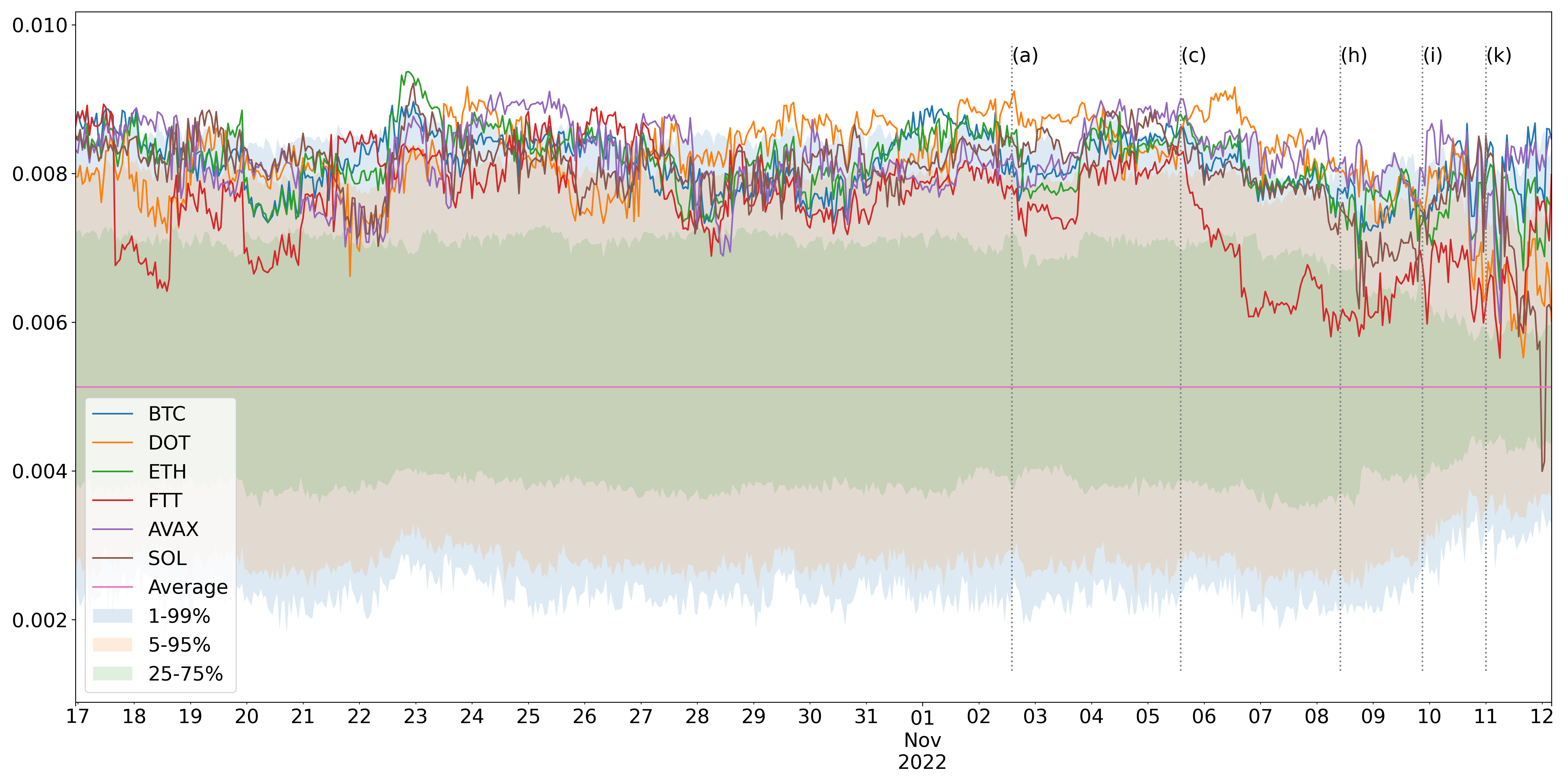}
    \includegraphics[width=\textwidth]{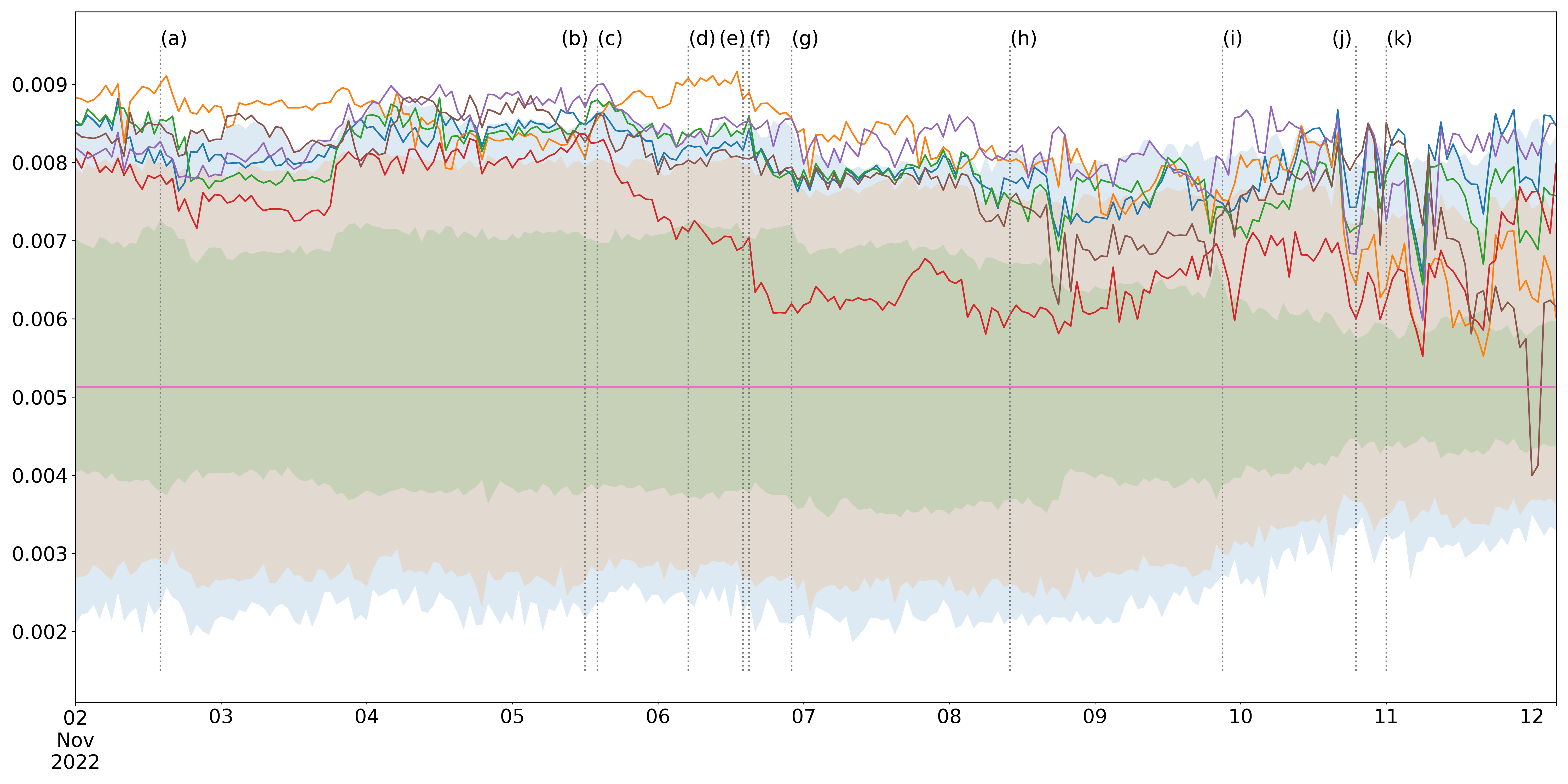}
    \caption{Rolling normalized closeness centrality on FTX market in the entire sample period (top) and the zoomed-in collapse period (bottom).}
    \label{fig:rolling_closenessN}
\end{figure}

\begin{figure}
    \centering
    \includegraphics[width=\textwidth]{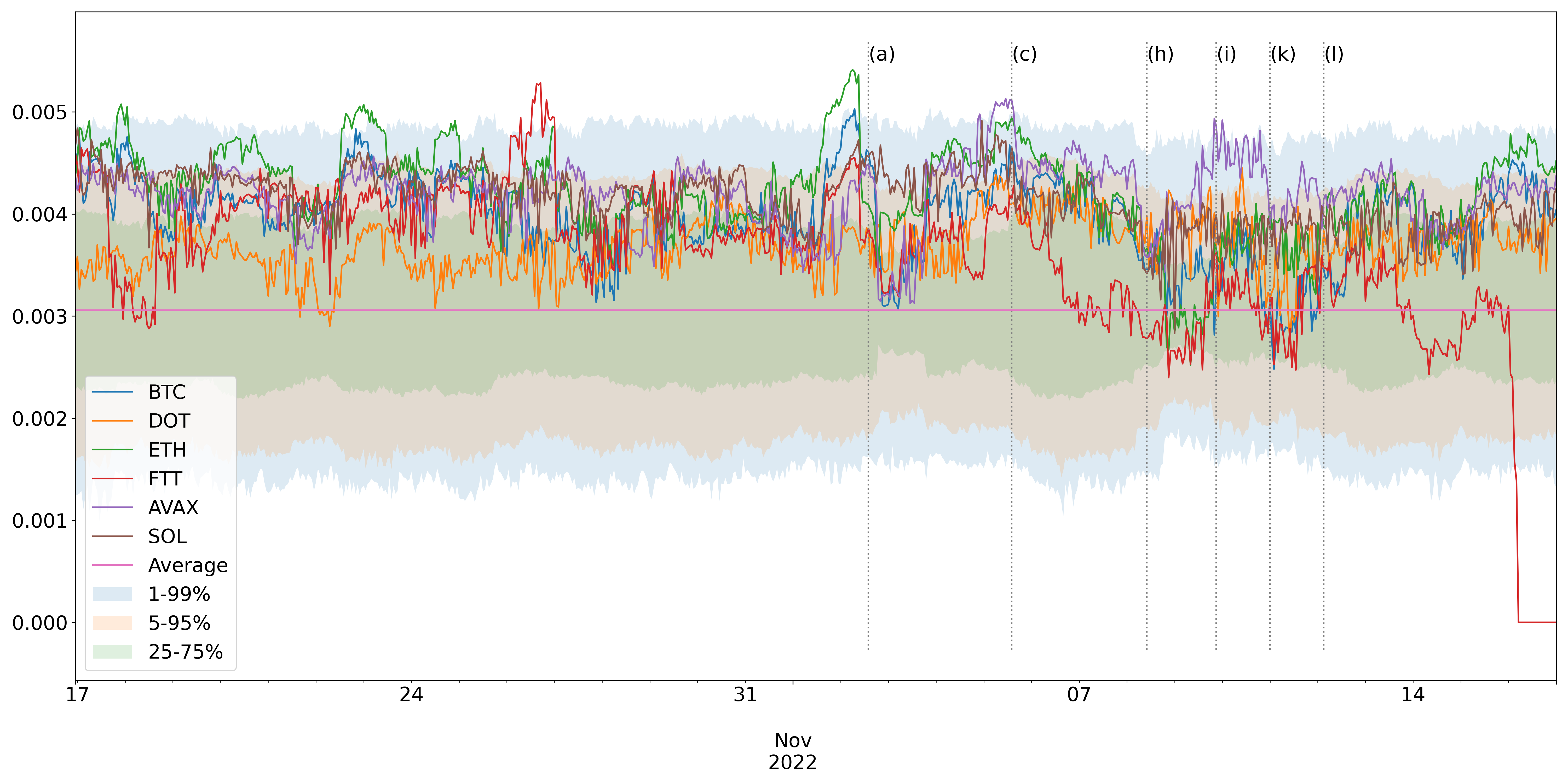}
    \includegraphics[width=\textwidth]{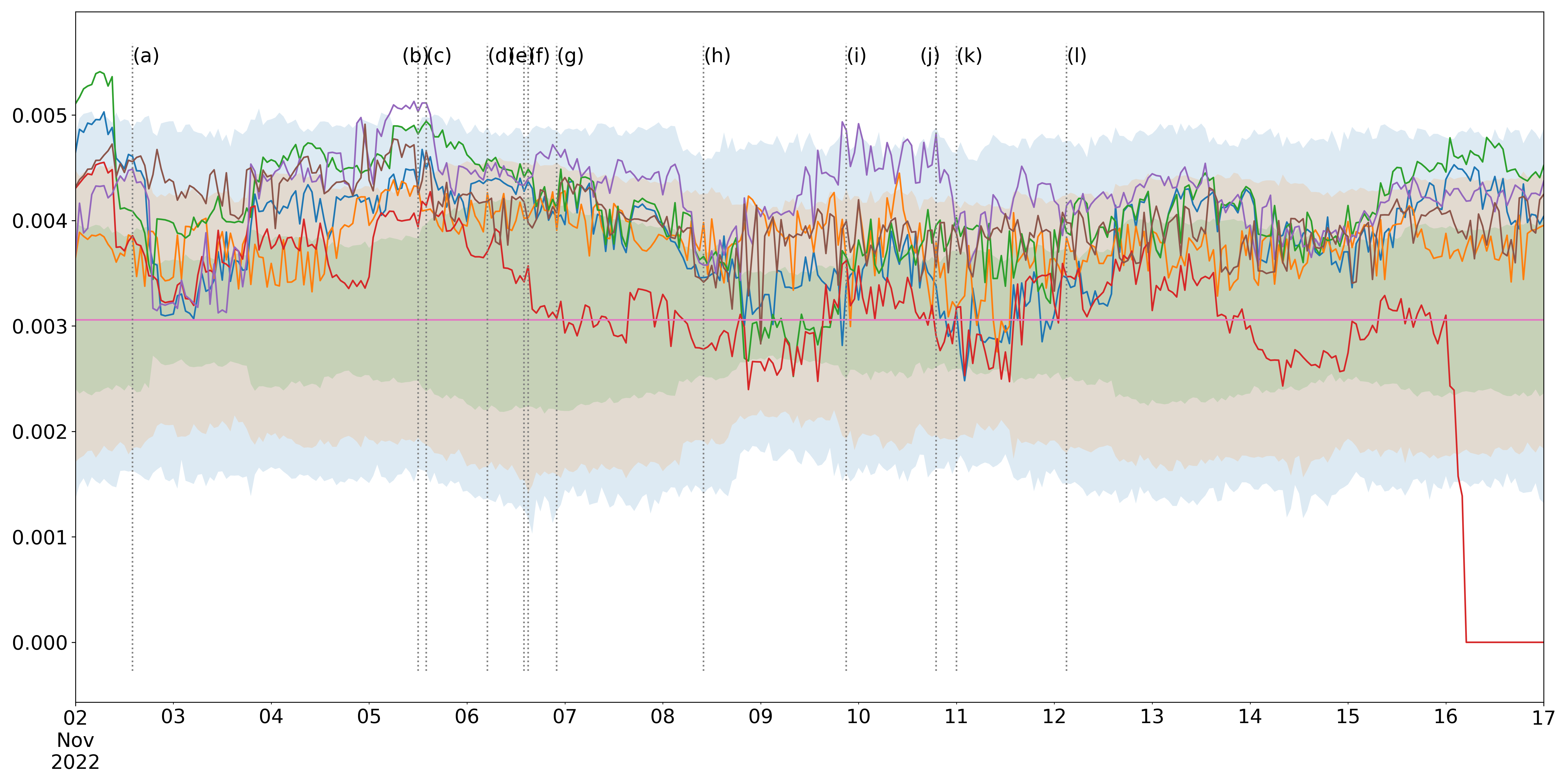}
    \caption{Rolling normalized closeness centrality on Binance market in the entire sample period (top) and the zoomed-in collapse period (bottom).}
    \label{fig:rolling_closenessNB}
\end{figure}

\clearpage

\section{List of cryptocurrencies}
\label{appendix_list}

\begin{table}
    \centering
    \resizebox*{!}{.9\textheight}{
    \begin{tabular}{lllll}
    1INCH & CHZ & GOG & MPLX & SNX \\
    AAVE & CITY & GRT & MSOL & SNY \\
    AGLD & CLV & GST & MTA & SOL \\
    AKRO & COMP & GT & MTL & SPA \\
    ALCX & CONV & HGET & MYC & SPELL \\
    ALEPH & COPE & HMT & NEAR & SRM \\
    ALGO & CQT & HNT & NEXO & STARS \\
    ALICE & CREAM & HOLY & OKB & STEP \\
    ALPHA & CRO & HT & OMG & STETH \\
    AMPL & CRV & HUM & OPENDAOSOS & STG \\
    APE & CVC & HXRO & ORBS & STMX \\
    APT & CVX & IMX & ORCA & STORJ \\
    ASD & DAI & INDI & OXY & STSOL \\
    ATLAS & DAWN & INTER & PAXG & SUN \\
    ATOM & DENT & IP3 & PEOPLE & SUSHI \\
    AUDIO & DFL & JOE & PERP & SWEAT \\
    AURY & DMG & JST & POLIS & SXP \\
    AVAX & DODO & KIN & PORT & TLM \\
    AXS & DOGE & KNC & PRISM & TOMO \\
    BADGER & DOT & KSHIB & PROM & TONCOIN \\
    BAL & DYDX & LDO & PSG & TRU \\
    BAND & EDEN & LEO & PSY & TRX \\
    BAO & EMB & LINA & PTU & TULIP \\
    BAR & ENJ & LINK & PUNDIX & UBXT \\
    BAT & ENS & LOOKS & QI & UMEE \\
    BCH & ETH & LRC & RAY & UNI \\
    BICO & ETHW & LTC & REALY & USDT \\
    BIT & EURT & LUA & REEF & VGX \\
    BLT & FIDA & MANA & REN & WAVES \\
    BNB & FRONT & MAPS & RNDR & WAXL \\
    BNT & FTM & MATH & ROOK & WFLOW \\
    BOBA & FTT & MATIC & RSR & WRX \\
    BTC & FXS & MBS & SAND & XAUT \\
    BTT & GAL & MCB & SECO & XPLA \\
    BVOL & GALA & MEDIA & SHIB & XRP \\
    C2X & GARI & MER & SKL & YFI \\
    C98 & GENE & MKR & SLND & YFII \\
    CEL & GMT & MNGO & SLP & YGG \\
    CHR & GODS & MOB & SLRS & ZRX \\
    \end{tabular}
    }
    \caption{List of the 195 cryptocurrencies (RIC codes) from the FTX market.}
    \label{tab:cryptolist}
\end{table}

\begin{table}
    \centering
    \resizebox*{!}{\textheight}{
    \begin{tabular}{llllll}
    1INCH & BSW & ERN & KP3R & ORN & STX \\
    AAVE & BTCST & ETC & KSM & OXT & SUN \\
    ACA & BTC & ETHERNITY & LAZIO & PAXG & SUSHI \\
    ACH & BTG & ETH & LDO & PEOPLE & SXP \\
    ACM & BTS & FARM & LEVER & PERL & SYS \\
    ADA & BTT & FET & LINA & PERP & TCT \\
    ADX & BURGER & FIDA & LINK & PHA & TFUEL \\
    AGLD & BUSD & FIL & LIT & PLA & THETA \\
    AION & C98 & FIO & LOKA & PNT & TKO \\
    AKRO & CAKE & FIRO & LPT & POLS & TLM \\
    ALCX & CELO & FIS & LRC & POLYX & TOMO \\
    ALGO & CELR & FLM & LSK & POND & TORN \\
    ALICE & CFX & FLOW & LTC & PORTO & TRB \\
    ALPACA & CHESS & FLUX & LTO & POWR & TRIBE \\
    ALPHA & CHR & FORTH & LUNA & PSG & TROY \\
    ALPINE & CHZ & FOR & LUNC & PUNDIX & TRU \\
    AMP & CITY & FRONT & MANA & PYR & TRX \\
    ANC & CKB & FTM & MASK & QI & T \\
    ANKR & CLV & FTT & MATIC & QNT & TVK \\
    ANT & COCOS & FUN & MBL & QTUM & TWT \\
    APE & COMP & FXS & MBOX & QUICK & UMA \\
    API3 & CONT & GALA & MC & RAD & UNFI \\
    APT & COTI & GAL & MDT & RARE & UNI \\
    ARDR & CRV & GHST & MDX & RAY & UTK \\
    ARPA & CTK & GLMR & MFT & REEF & VET \\
    AR & CTSI & GMT & MINA & REI & VGX \\
    ASR & CTXC & GNO & MIOTA & REN & VIDT \\
    ASTR & CVC & GRT & MIR & REP & VITE \\
    ATA & CVP & GTC & MITH & REQ & VOXEL \\
    ATM & CVX & GTO & MKR & RIF & VTHO \\
    ATOM & DAR & HARD & MLN & RLC & WAN \\
    AUCTION & DASH & HBAR & MOB & RNDR & WAVES \\
    AUDIO & DATA & HIGH & MOVR & ROSE & WAXP \\
    AUTO & DCR & HIVE & MTL & RSR & WING \\
    AVA & DEGO & HOT & MULTI & RUNE & WIN \\
    AVAX & DENT & ICP & NANO & RVN & WNXM \\
    AXS & DEXE & ICX & NBS & SAND & WOO \\
    BADGER & DF & IDEX & NBT & SANTOS & WRX \\
    BAKE & DGB & ILV & NEAR & SCRT & WTC \\
    BAL & DIA & IMX & NEO & SC & XEC \\
    BAND & DNT & INJ & NEXO & SFP & XEM \\
    BAR & DOCK & IOST & NKN & SHIB & XLM \\
    BAT & DODO & IOTX & NMR & SKL & XMR \\
    BCH & DOGE & IRIS & NULS & SLP & XRP \\
    BEAM & DOT & JASMY & OCEAN & SNX & XTZ \\
    BEL & DREP & JOE & OGN & SOL & XVG \\
    BETA & DUSK & JST & OG & SPELL & XVS \\
    BICO & DYDX & JUV & OMG & SRM & YFII \\
    BIFI & EGLD & KAVA & OM & STEEM & YFI \\
    BLZ & ELF & KDA & ONE & STG & YGG \\
    BNB & ENJ & KEY & ONGAS & STMX & ZEC \\
    BNT & ENS & KLAY & ONT & STORJ & ZEN \\
    BNX & EOS & KMD & OOKI & STPT & ZIL \\
    BOND & EPX & KNC & OP & STRAX & ZRX \\
    \end{tabular}
    }
    \caption{List of the 324 cryptocurrencies (RIC codes) from the Binance market.}
    \label{tab:my_label}
\end{table}

\clearpage
  \bibliographystyle{elsarticle-num-names} 
  \bibliography{main}

\begin{thebibliography}{43}
\expandafter\ifx\csname natexlab\endcsname\relax\def\natexlab#1{#1}\fi
\providecommand{\url}[1]{\texttt{#1}}
\providecommand{\href}[2]{#2}
\providecommand{\path}[1]{#1}
\providecommand{\DOIprefix}{doi:}
\providecommand{\ArXivprefix}{arXiv:}
\providecommand{\URLprefix}{URL: }
\providecommand{\Pubmedprefix}{pmid:}
\providecommand{\doi}[1]{\href{http://dx.doi.org/#1}{\path{#1}}}
\providecommand{\Pubmed}[1]{\href{pmid:#1}{\path{#1}}}
\providecommand{\bibinfo}[2]{#2}
\ifx\xfnm\relax \def\xfnm[#1]{\unskip,\space#1}\fi
\bibitem[{Guo and Donev(2020)}]{guo2020}
\bibinfo{author}{X.~Guo}, \bibinfo{author}{P.~Donev},
\newblock \bibinfo{title}{Bibliometrics and network analysis of cryptocurrency
  research},
\newblock \bibinfo{journal}{Journal of {S}ystems {S}cience and {C}omplexity}
  \bibinfo{volume}{33} (\bibinfo{year}{2020}) \bibinfo{pages}{1933--1958}.
\bibitem[{Fang et~al.(2022)Fang, Ventre, Basios, Kanthan, {Martinez-Rego}, Wu,
  and Li}]{fang2022}
\bibinfo{author}{F.~Fang}, \bibinfo{author}{C.~Ventre},
  \bibinfo{author}{M.~Basios}, \bibinfo{author}{L.~Kanthan},
  \bibinfo{author}{D.~{Martinez-Rego}}, \bibinfo{author}{F.~Wu},
  \bibinfo{author}{L.~Li},
\newblock \bibinfo{title}{Cryptocurrency trading: A comprehensive survey},
\newblock \bibinfo{journal}{Financial Innovation} \bibinfo{volume}{8}
  (\bibinfo{year}{2022}) \bibinfo{pages}{13}.
  \DOIprefix\doi{10.1186/s40854-021-00321-6}.
\bibitem[{Wu et~al.(2021)Wu, Liu, Zhao, and Zheng}]{wu2021}
\bibinfo{author}{J.~Wu}, \bibinfo{author}{J.~Liu}, \bibinfo{author}{Y.~Zhao},
  \bibinfo{author}{Z.~Zheng},
\newblock \bibinfo{title}{Analysis of cryptocurrency transactions from a
  network perspective: {{An}} overview},
\newblock \bibinfo{journal}{Journal of Network and Computer Applications}
  \bibinfo{volume}{190} (\bibinfo{year}{2021}) \bibinfo{pages}{103139}.
  \DOIprefix\doi{10.1016/j.jnca.2021.103139}.
\bibitem[{Zou et~al.(2019)Zou, Donner, Marwan, Donges, and Kurths}]{zou2019}
\bibinfo{author}{Y.~Zou}, \bibinfo{author}{R.~V. Donner},
  \bibinfo{author}{N.~Marwan}, \bibinfo{author}{J.~F. Donges},
  \bibinfo{author}{J.~Kurths},
\newblock \bibinfo{title}{Complex network approaches to nonlinear time series
  analysis},
\newblock \bibinfo{journal}{Physics {R}eports} \bibinfo{volume}{787}
  (\bibinfo{year}{2019}) \bibinfo{pages}{1--97}.
\bibitem[{Mantegna(1999)}]{mantegna1999}
\bibinfo{author}{R.~N. Mantegna},
\newblock \bibinfo{title}{Hierarchical structure in financial markets},
\newblock \bibinfo{journal}{The European {P}hysical {J}ournal {B}-{C}ondensed
  {M}atter and {C}omplex {S}ystems} \bibinfo{volume}{11} (\bibinfo{year}{1999})
  \bibinfo{pages}{193--197}.
\bibitem[{Onnela et~al.(2003)Onnela, Chakraborti, Kaski, Kert{\'e}sz, and
  Kanto}]{onnela2003}
\bibinfo{author}{J.-P. Onnela}, \bibinfo{author}{A.~Chakraborti},
  \bibinfo{author}{K.~Kaski}, \bibinfo{author}{J.~Kert{\'e}sz},
  \bibinfo{author}{A.~Kanto},
\newblock \bibinfo{title}{Asset {{Trees}} and {{Asset Graphs}} in {{Financial
  Markets}}},
\newblock \bibinfo{journal}{Physica Scripta} \bibinfo{volume}{2003}
  (\bibinfo{year}{2003}) \bibinfo{pages}{48}.
  \DOIprefix\doi{10.1238/Physica.Topical.106a00048}.
\bibitem[{Tumminello et~al.(2005)Tumminello, Aste, Di~Matteo, and
  Mantegna}]{tumminello2005}
\bibinfo{author}{M.~Tumminello}, \bibinfo{author}{T.~Aste},
  \bibinfo{author}{T.~Di~Matteo}, \bibinfo{author}{R.~N. Mantegna},
\newblock \bibinfo{title}{A tool for filtering information in complex systems},
\newblock \bibinfo{journal}{Proceedings of the {N}ational {A}cademy of
  {S}ciences} \bibinfo{volume}{102} (\bibinfo{year}{2005})
  \bibinfo{pages}{10421--10426}.
\bibitem[{Tumminello et~al.(2007)Tumminello, Di~Matteo, Aste, and
  Mantegna}]{tumminello2007}
\bibinfo{author}{M.~Tumminello}, \bibinfo{author}{T.~Di~Matteo},
  \bibinfo{author}{T.~Aste}, \bibinfo{author}{R.~N. Mantegna},
\newblock \bibinfo{title}{Correlation based networks of equity returns sampled
  at different time horizons},
\newblock \bibinfo{journal}{The {E}uropean {P}hysical {J}ournal {B}}
  \bibinfo{volume}{55} (\bibinfo{year}{2007}) \bibinfo{pages}{209--217}.
\bibitem[{Massara et~al.(2017)Massara, Di~Matteo, and Aste}]{massara2016}
\bibinfo{author}{G.~P. Massara}, \bibinfo{author}{T.~Di~Matteo},
  \bibinfo{author}{T.~Aste},
\newblock \bibinfo{title}{Network filtering for big data: {T}riangulated
  maximally filtered graph},
\newblock \bibinfo{journal}{Journal of {C}omplex {N}etworks}
  \bibinfo{volume}{5} (\bibinfo{year}{2017}) \bibinfo{pages}{161--178}.
\bibitem[{Pilar et~al.(2018)Pilar, Jaureguizar~Arellano, and
  Jaureguizar~Franc{\'e}s}]{pilar2018}
\bibinfo{author}{G.-C. Pilar}, \bibinfo{author}{D.~Jaureguizar~Arellano},
  \bibinfo{author}{C.~Jaureguizar~Franc{\'e}s},
\newblock \bibinfo{title}{The cryptocurrency market: {{A}} network analysis},
\newblock \bibinfo{journal}{ESIC MARKET Economic and Business Journal}
  \bibinfo{volume}{49} (\bibinfo{year}{2018}).
  \DOIprefix\doi{10.7200/esicm.161.0493.4i}.
\bibitem[{Scagliarini et~al.(2022)Scagliarini, Pappalardo, Biondo, Pluchino,
  Rapisarda, and Stramaglia}]{scagliarini2022}
\bibinfo{author}{T.~Scagliarini}, \bibinfo{author}{G.~Pappalardo},
  \bibinfo{author}{A.~E. Biondo}, \bibinfo{author}{A.~Pluchino},
  \bibinfo{author}{A.~Rapisarda}, \bibinfo{author}{S.~Stramaglia},
\newblock \bibinfo{title}{Pairwise and high-order dependencies in the
  cryptocurrency trading network},
\newblock \bibinfo{journal}{Scientific Reports} \bibinfo{volume}{12}
  (\bibinfo{year}{2022}) \bibinfo{pages}{18483}.
  \DOIprefix\doi{10.1038/s41598-022-21192-6}.
\bibitem[{Granger(1969)}]{granger1969}
\bibinfo{author}{C.~W. Granger},
\newblock \bibinfo{title}{Investigating causal relations by econometric models
  and cross-spectral methods},
\newblock \bibinfo{journal}{Econometrica: {J}ournal of the Econometric Society}
   (\bibinfo{year}{1969}) \bibinfo{pages}{424--438}.
\bibitem[{Nie(2022)}]{nie2022}
\bibinfo{author}{C.-X. Nie},
\newblock \bibinfo{title}{Analysis of critical events in the correlation
  dynamics of cryptocurrency market},
\newblock \bibinfo{journal}{Physica A: Statistical Mechanics and its
  Applications} \bibinfo{volume}{586} (\bibinfo{year}{2022})
  \bibinfo{pages}{126462}. \DOIprefix\doi{10.1016/j.physa.2021.126462}.
\bibitem[{Liao et~al.(2024)Liao, Li, Chan, Chu, and
  Zhang}]{liao_interconnections_2024}
\bibinfo{author}{X.~Liao}, \bibinfo{author}{Q.~Li}, \bibinfo{author}{S.~Chan},
  \bibinfo{author}{J.~Chu}, \bibinfo{author}{Y.~Zhang},
\newblock \bibinfo{title}{Interconnections and contagion among
  cryptocurrencies, {DeFi}, {NFT} and traditional financial assets: {Some} new
  evidence from tail risk driven network},
\newblock \bibinfo{journal}{Physica A: Statistical Mechanics and its
  Applications} \bibinfo{volume}{647} (\bibinfo{year}{2024})
  \bibinfo{pages}{129892}. \DOIprefix\doi{10.1016/j.physa.2024.129892}.
\bibitem[{Galati et~al.(2024)Galati, Webb, and Webb}]{galati_market_2024}
\bibinfo{author}{L.~Galati}, \bibinfo{author}{A.~Webb}, \bibinfo{author}{R.~I.
  Webb},
\newblock \bibinfo{title}{Market behaviors around bankruptcy and frozen funds
  withdrawal: {Trading} stranded assets on {FTX}},
\newblock \bibinfo{journal}{Journal of Economics and Business}
  (\bibinfo{year}{2024}) \bibinfo{pages}{106196}.
  \DOIprefix\doi{10.1016/j.jeconbus.2024.106196}.
\bibitem[{De~Blasis et~al.(2023)De~Blasis, Galati, Webb, and
  Webb}]{de_blasis_intelligent_2023}
\bibinfo{author}{R.~De~Blasis}, \bibinfo{author}{L.~Galati},
  \bibinfo{author}{A.~Webb}, \bibinfo{author}{R.~I. Webb},
\newblock \bibinfo{title}{Intelligent design: stablecoins (in)stability and
  collateral during market turbulence},
\newblock \bibinfo{journal}{Financial Innovation} \bibinfo{volume}{9}
  (\bibinfo{year}{2023}) \bibinfo{pages}{85}.
  \DOIprefix\doi{10.1186/s40854-023-00492-4}.
\bibitem[{Galati and Capalbo(2024)}]{galati_silicon_2024}
\bibinfo{author}{L.~Galati}, \bibinfo{author}{F.~Capalbo},
\newblock \bibinfo{title}{Silicon {Valley} {Bank} bankruptcy and {Stablecoins}
  stability},
\newblock \bibinfo{journal}{International Review of Financial Analysis}
  \bibinfo{volume}{91} (\bibinfo{year}{2024}) \bibinfo{pages}{103001}.
  \DOIprefix\doi{10.1016/j.irfa.2023.103001}.
\bibitem[{Galati et~al.(2024)Galati, Webb, and Webb}]{galati_financial_2024}
\bibinfo{author}{L.~Galati}, \bibinfo{author}{A.~Webb}, \bibinfo{author}{R.~I.
  Webb},
\newblock \bibinfo{title}{Financial contagion in cryptocurrency exchanges:
  {Evidence} from the {FTT} collapse},
\newblock \bibinfo{journal}{Finance Research Letters}  (\bibinfo{year}{2024})
  \bibinfo{pages}{105747}. \DOIprefix\doi{10.1016/j.frl.2024.105747}.
\bibitem[{Esparcia et~al.(2024)Esparcia, Escribano, and
  Jareño}]{esparcia_assessing_2024}
\bibinfo{author}{C.~Esparcia}, \bibinfo{author}{A.~Escribano},
  \bibinfo{author}{F.~Jareño},
\newblock \bibinfo{title}{Assessing the crypto market stability after the {FTX}
  collapse: {A} study of high frequency volatility and connectedness},
\newblock \bibinfo{journal}{International Review of Financial Analysis}
  \bibinfo{volume}{94} (\bibinfo{year}{2024}) \bibinfo{pages}{103287}.
  \DOIprefix\doi{10.1016/j.irfa.2024.103287}.
\bibitem[{Jalan and Matkovskyy(2023)}]{jalan_systemic_2023}
\bibinfo{author}{A.~Jalan}, \bibinfo{author}{R.~Matkovskyy},
\newblock \bibinfo{title}{Systemic risks in the cryptocurrency market:
  {Evidence} from the {FTX} collapse},
\newblock \bibinfo{journal}{Finance Research Letters} \bibinfo{volume}{53}
  (\bibinfo{year}{2023}) \bibinfo{pages}{103670}.
  \DOIprefix\doi{10.1016/j.frl.2023.103670}.
\bibitem[{Akyildirim et~al.(2023)Akyildirim, Conlon, Corbet, and
  Goodell}]{akyildirim_understanding_2023}
\bibinfo{author}{E.~Akyildirim}, \bibinfo{author}{T.~Conlon},
  \bibinfo{author}{S.~Corbet}, \bibinfo{author}{J.~W. Goodell},
\newblock \bibinfo{title}{Understanding the {FTX} exchange collapse: {A}
  dynamic connectedness approach},
\newblock \bibinfo{journal}{Finance Research Letters} \bibinfo{volume}{53}
  (\bibinfo{year}{2023}) \bibinfo{pages}{103643}.
  \DOIprefix\doi{10.1016/j.frl.2023.103643}.
\bibitem[{Yousaf and Goodell(2023)}]{yousaf_reputational_2023}
\bibinfo{author}{I.~Yousaf}, \bibinfo{author}{J.~W. Goodell},
\newblock \bibinfo{title}{Reputational contagion and the fall of {FTX}:
  {Examining} the response of tokens to the delegitimization of {FTT}},
\newblock \bibinfo{journal}{Finance Research Letters} \bibinfo{volume}{54}
  (\bibinfo{year}{2023}) \bibinfo{pages}{103704}.
  \DOIprefix\doi{10.1016/j.frl.2023.103704}.
\bibitem[{Yousaf et~al.(2023)Yousaf, Riaz, and Goodell}]{yousaf_what_2023}
\bibinfo{author}{I.~Yousaf}, \bibinfo{author}{Y.~Riaz}, \bibinfo{author}{J.~W.
  Goodell},
\newblock \bibinfo{title}{What do responses of financial markets to the
  collapse of {FTX} say about investor interest in cryptocurrencies?
  {Event}-study evidence},
\newblock \bibinfo{journal}{Finance Research Letters} \bibinfo{volume}{53}
  (\bibinfo{year}{2023}) \bibinfo{pages}{103661}.
  \DOIprefix\doi{10.1016/j.frl.2023.103661}.
\bibitem[{Conlon et~al.(2023)Conlon, Corbet, and Hu}]{conlon_collapse_2023}
\bibinfo{author}{T.~Conlon}, \bibinfo{author}{S.~Corbet},
  \bibinfo{author}{Y.~Hu},
\newblock \bibinfo{title}{The collapse of the {FTX} exchange: {The} end of
  cryptocurrency's age of innocence},
\newblock \bibinfo{journal}{The British Accounting Review}
  (\bibinfo{year}{2023}) \bibinfo{pages}{101277}.
  \DOIprefix\doi{10.1016/j.bar.2023.101277}.
\bibitem[{Conlon et~al.(2024)Conlon, Corbet, and Hou}]{conlon_contagion_2024}
\bibinfo{author}{T.~Conlon}, \bibinfo{author}{S.~Corbet},
  \bibinfo{author}{Y.~G. Hou},
\newblock \bibinfo{title}{Contagion effects of permissionless, worthless
  cryptocurrency tokens: {Evidence} from the collapse of {FTX}},
\newblock \bibinfo{journal}{Journal of International Financial Markets,
  Institutions and Money} \bibinfo{volume}{91} (\bibinfo{year}{2024})
  \bibinfo{pages}{101940}. \DOIprefix\doi{10.1016/j.intfin.2024.101940}.
\bibitem[{Wang et~al.(2024)Wang, Lu, Lin, Ren, and H{\"a}rdle}]{wang2024}
\bibinfo{author}{Y.~Wang}, \bibinfo{author}{W.~Lu}, \bibinfo{author}{M.-B.
  Lin}, \bibinfo{author}{R.~Ren}, \bibinfo{author}{W.~K. H{\"a}rdle},
\newblock \bibinfo{title}{Cross-exchange crypto risk: {{A}} high-frequency
  dynamic network perspective},
\newblock \bibinfo{journal}{International Review of Financial Analysis}
  \bibinfo{volume}{94} (\bibinfo{year}{2024}) \bibinfo{pages}{103246}.
  \DOIprefix\doi{10.1016/j.irfa.2024.103246}.
\bibitem[{Vidal-Tomás et~al.(2023)Vidal-Tomás, Briola, and
  Aste}]{vidal-tomas_ftxs_2023}
\bibinfo{author}{D.~Vidal-Tomás}, \bibinfo{author}{A.~Briola},
  \bibinfo{author}{T.~Aste},
\newblock \bibinfo{title}{{FTX}’s downfall and {Binance}’s consolidation:
  {The} fragility of centralised digital finance},
\newblock \bibinfo{journal}{Physica A: Statistical Mechanics and its
  Applications} \bibinfo{volume}{625} (\bibinfo{year}{2023})
  \bibinfo{pages}{129044}. \DOIprefix\doi{10.1016/j.physa.2023.129044}.
\bibitem[{Galati(2024)}]{galati_exchange_2024}
\bibinfo{author}{L.~Galati},
\newblock \bibinfo{title}{Exchange market share, market makers, and murky
  behavior: {The} impact of no-fee trading on cryptocurrency market quality},
\newblock \bibinfo{journal}{Journal of Banking \& Finance}
  \bibinfo{volume}{165} (\bibinfo{year}{2024}) \bibinfo{pages}{107222}.
  \DOIprefix\doi{10.1016/j.jbankfin.2024.107222}.
\bibitem[{Harary(1969)}]{Harary}
\bibinfo{author}{F.~Harary}, \bibinfo{title}{{Graph theory}},
  \bibinfo{publisher}{Addison-Wesley}, \bibinfo{year}{1969}.
\bibitem[{Estrada(2012)}]{Estrada2012}
\bibinfo{author}{E.~Estrada}, \bibinfo{title}{The structure of complex
  networks: theory and applications}, \bibinfo{publisher}{Oxford University
  Press}, \bibinfo{year}{2012}.
\bibitem[{West et~al.(2001)}]{west2001}
\bibinfo{author}{D.~B. West}, et~al., \bibinfo{title}{Introduction to graph
  theory}, volume~\bibinfo{volume}{2}, \bibinfo{publisher}{Prentice hall
  {U}pper {S}addle {R}iver}, \bibinfo{year}{2001}.
\bibitem[{Nishizeki and Chiba(2008)}]{Nishizeki2008}
\bibinfo{author}{T.~Nishizeki}, \bibinfo{author}{N.~Chiba},
  \bibinfo{title}{Planar graphs: Theory and algorithms},
  \bibinfo{publisher}{Courier Corporation}, \bibinfo{year}{2008}.
\bibitem[{Aste et~al.(2005)Aste, Di~Matteo, and Hyde}]{aste2005}
\bibinfo{author}{T.~Aste}, \bibinfo{author}{T.~Di~Matteo},
  \bibinfo{author}{S.~Hyde},
\newblock \bibinfo{title}{Complex networks on hyperbolic surfaces},
\newblock \bibinfo{journal}{Physica {A}: {S}tatistical {M}echanics and its
  {A}pplications} \bibinfo{volume}{346} (\bibinfo{year}{2005})
  \bibinfo{pages}{20--26}.
\bibitem[{Freeman(1978)}]{Freeman1978}
\bibinfo{author}{L.~C. Freeman},
\newblock \bibinfo{title}{Centrality in social networks conceptual
  clarification},
\newblock \bibinfo{journal}{Social Networks} \bibinfo{volume}{1}
  (\bibinfo{year}{1978}) \bibinfo{pages}{215--239}.
\bibitem[{Opsahl et~al.(2010)Opsahl, Agneessens, and Skvoretz}]{Opsahl2010}
\bibinfo{author}{T.~Opsahl}, \bibinfo{author}{F.~Agneessens},
  \bibinfo{author}{J.~Skvoretz},
\newblock \bibinfo{title}{Node centrality in weighted networks: Generalizing
  degree and shortest paths},
\newblock \bibinfo{journal}{Social networks} \bibinfo{volume}{32}
  (\bibinfo{year}{2010}) \bibinfo{pages}{245--251}.
\bibitem[{Schrijver(2012)}]{schrijver2012}
\bibinfo{author}{A.~Schrijver},
\newblock \bibinfo{title}{On the history of the shortest path problem},
\newblock \bibinfo{journal}{Documenta {M}athematica} \bibinfo{volume}{17}
  (\bibinfo{year}{2012}) \bibinfo{pages}{155--167}.
\bibitem[{Dijkstra(1959)}]{Dijkstra1959}
\bibinfo{author}{E.~Dijkstra},
\newblock \bibinfo{title}{A note on two problems in connexion with graphs},
\newblock \bibinfo{journal}{Numerische {M}athematik}  (\bibinfo{year}{1959})
  \bibinfo{pages}{269–271}.
  \DOIprefix\doi{https://doi.org/10.1007/BF01386390}.
\bibitem[{Newman(2001)}]{newman2001}
\bibinfo{author}{M.~E. Newman},
\newblock \bibinfo{title}{Scientific collaboration networks. ii. shortest
  paths, weighted networks, and centrality},
\newblock \bibinfo{journal}{Physical {R}eview {E}} \bibinfo{volume}{64}
  (\bibinfo{year}{2001}) \bibinfo{pages}{016132}.
\bibitem[{Brandes(2001)}]{brandes2001}
\bibinfo{author}{U.~Brandes},
\newblock \bibinfo{title}{A faster algorithm for betweenness centrality},
\newblock \bibinfo{journal}{Journal of mathematical sociology}
  \bibinfo{volume}{25} (\bibinfo{year}{2001}) \bibinfo{pages}{163--177}.
\bibitem[{Latora and Marchiori(2001)}]{latora2001}
\bibinfo{author}{V.~Latora}, \bibinfo{author}{M.~Marchiori},
\newblock \bibinfo{title}{Efficient behavior of small-world networks},
\newblock \bibinfo{journal}{Physical {R}eview {L}etters} \bibinfo{volume}{87}
  (\bibinfo{year}{2001}) \bibinfo{pages}{198701}.
\bibitem[{Latora and Marchiori(2007)}]{latora2007}
\bibinfo{author}{V.~Latora}, \bibinfo{author}{M.~Marchiori},
\newblock \bibinfo{title}{A measure of centrality based on network efficiency},
\newblock \bibinfo{journal}{New {J}ournal of {P}hysics} \bibinfo{volume}{9}
  (\bibinfo{year}{2007}) \bibinfo{pages}{188}.
\bibitem[{Shirokikh et~al.(2013)Shirokikh, Pastukhov, Boginski, and
  Butenko}]{Shirokikh2013}
\bibinfo{author}{O.~Shirokikh}, \bibinfo{author}{G.~Pastukhov},
  \bibinfo{author}{V.~Boginski}, \bibinfo{author}{S.~Butenko},
\newblock \bibinfo{title}{Computational study of the us stock market evolution:
  a rank correlation-based network model},
\newblock \bibinfo{journal}{Computational Management Science}
  \bibinfo{volume}{10} (\bibinfo{year}{2013}) \bibinfo{pages}{81--103}.
\bibitem[{Chen et~al.(2021)Chen, Qu, Jiang, and Jiang}]{Chen2021}
\bibinfo{author}{W.~Chen}, \bibinfo{author}{S.~Qu}, \bibinfo{author}{M.~Jiang},
  \bibinfo{author}{C.~Jiang},
\newblock \bibinfo{title}{The construction of multilayer stock network model},
\newblock \bibinfo{journal}{Physica A: Statistical Mechanics and its
  Applications} \bibinfo{volume}{565} (\bibinfo{year}{2021})
  \bibinfo{pages}{125608}.

\end{thebibliography}






\end{document}